\documentclass[prb, twocolumn]{revtex4}

\begin{document}

\title{
Topological properties of Abelian and non-Abelian quantum Hall states\\
from the pattern of zeros
}

\author{Xiao-Gang Wen}
\affiliation{
Department of Physics,
Massachusetts Institute of Technology,
Cambridge, MA 02139, USA
}

\author{Zhenghan Wang}
\affiliation{
Microsoft Station Q, CNSI Bldg. Rm 2237, University
of California, Santa Barbara, CA 93106 }

\date{March 6, 2008}

\begin{abstract}
It has been shown that different Abelian and non-Abelian fraction quantum Hall
states can be characterized by patterns of zeros described by sequences of
integers $\{S_a\}$.  In this paper, we will show how to use the data $\{S_a\}$
to calculate various topological properties of the corresponding fraction
quantum Hall state, such as the number of possible quasiparticle types and
their quantum numbers, as well as the actions of the quasiparticle tunneling
and modular transformations on the degenerate ground states on torus.
\end{abstract}

\maketitle

\setcounter{tocdepth}{3}
\tableofcontents

\section{Introduction}

Characterization and classification of wave functions with infinite variables
are the key to gain a deeper understanding of quantum phases and quantum phase
transitions.  For a long time, people have used symmetry properties of the
wave functions and the change of those symmetries to understand  quantum
phases and quantum phase transitions.  However, the discovery of fractional
quantum Hall (FQH) states \cite{TSG8259,L8395} suggests that the symmetry
characterization is not enough to describe different FQH states. Thus FQH
states contain a new kind of order -- topological order.\cite{WNtop,Wtoprev}
Topological order is so new that the conventional mathematical tools and the
language developed for symmetry breaking orders are not adequate to describe
it.  Thus, looking for new mathematical framework to describe the topological
order becomes an important task in advancing the theory of topological
order.\cite{FNS0428,LWstrnet,K062,BK} Intuitively, topological order
correspond to pattern of long-range quantum entanglement in the ground state,
as revealed by the topological entanglement entropy.\cite{KP0604,LWtopent}
Characterizing topological order is characterizing patterns of long-range
entanglement.

In a recent paper,\cite{WWsymm} the pattern of zeros is introduced to
characterize and classify symmetric polynomials of infinite
variables. Applying this result to FQH states, we find that the
pattern of zeros characterizes and classifies many Abelian and
non-Abelian FQH states.  In other words, the pattern of zeros
characterizes the topological order in FQH states.\cite{Wtoprev}
This may lead to a deeper understanding of topological order in
FQH states.

More concretely, a pattern of zeros that characterizes FQH ground state wave
functions is described a sequence of integers $\{S_a\}$.  Describing pattern of
zeros through a sequence of integers $\{S_a\}$ generalizes the pseudo potential
construction of ideal Hamiltonian and the associated zero-energy ground
state.\cite{H8305,GWW9267,W9355,RR9984,R0634,SRC0760} All the topological
properties of FQH states should be determined by the data $\{S_a\}$.  In this
paper, we will show how to use the data $\{S_a\}$ to calculate those
topological properties. They include the quantum number of possible
quasiparticles and the ground state degeneracy on torus.\cite{H8595,Wtop}  We
also study the actions of quasiparticle tunneling and the modular
transformation on the degenerate ground states.\cite{Wrig} The main results of
the paper are summarized in Section \ref{exmpl}.

\section{Fusion and pattern of zeros}

In \Ref{WWsymm}, we have introduced the pattern of zeros of a wave function by
bringing $a$ variables together.  In this section,  we will review the pattern
of zeros within such an approach, but from a slightly different angle.  For
readers who are interested in the new results can go directly to section
\ref{qp}.

\subsection{Fusion of $a$ variables}
\label{secSa}

To bring $a$ variables in a symmetric polynomial $\Phi(\{z_i\})$ together, let
us view $z_1, \cdots,z_a$ as variables and
$z_{a+1},z_{a+2},\cdots$ as fixed parameters.  We write
$\Phi(z_1,z_2,\cdots)$ as
\begin{equation*}
 \Phi(\{z_i\})=\Phi(z_1,\cdots,z_a;z_{a+1},z_{a+2},\cdots)
\end{equation*}
where $\Phi(z_1,\cdots,z_a;z_{a+1},z_{a+2},\cdots)$ is a symmetric polynomial
of $z_1,\cdots,z_a$ parameterized by $(z_{a+1},z_{a+2},\cdots)$.
Now let us rewrite
\begin{align*}
 z_i&=\la \xi_i+z^{(a)},\ \ \ z^{(a)}=\frac{z_1+\cdots+z_a}{a},
\nonumber\\
& \sum_{i=1}^a \xi_i=0, \ \ \ i=1,\cdots,a ,
\end{align*}
and expand $\Phi(\la \xi_1+z^{(a)},\cdots,\la \xi_a+z^{(a)};
z_{a+1},z_{a+2},\cdots)$ in powers of $\la$:
\begin{align*}
&\ \ \  \Phi(\la \xi_1+z^{(a)},\cdots,\la \xi_a+z^{(a)};z_{a+1},z_{a+2},\cdots)
\nonumber\\
&= \sum_{m=0}^{\infty}\la^{m} P_m[z^{(a)},(\xi_1,\cdots,\xi_a);z_{a+1},z_{a+2},\cdots]
\end{align*}
where $P_m[z^{(a)},(\xi_1,\cdots,\xi_a);z_{a+1},z_{a+2},\cdots]$ is a
homogenous symmetric polynomial of $\xi_1,\cdots,\xi_a$ of degree $m$.
$P_m[z^{(a)},(\xi_1,\cdots,\xi_a);z_{a+1},z_{a+2},\cdots]$ is also a
polynomial of $z^{(a)}$ and $z_{a+1},z_{a+2},\cdots$. The minimal power of
$z^{(a)}$ is zero.

The polynomial
$P_m[z^{(a)},(\xi_1,\cdots,\xi_a);z_{a+1},z_{a+2},\cdots]$ is
derived from $\Phi(\{z_i\})$ after fusing $a$ $z_i$-variables into
a single $z^{(a)}$ variable.  Different choices of
$\xi_1,\cdots,\xi_a$ represent different ways to fuse
$z_1,\cdots,z_a$ into $z^{(a)}$.  As a polynomial of
$z^{(a)},z_{a+1},z_{a+2},\cdots$, different ways of fusion can
produce linearly independent
$P_m[z^{(a)},(\xi_1,\cdots,\xi_a);z_{a+1},z_{a+2},\cdots]$. Thus
$(\xi_1,\cdots,\xi_a)$ can be viewed as a label for those linearly
independent polynomials of $z^{(a)},z_{a+1},z_{a+2},\cdots$.

$P_m[z^{(a)},(\xi_1,\cdots,\xi_a);z_{a+1},z_{a+2},\cdots]$ is also a
symmetric polynomial of $\xi_1,\cdots,\xi_a$.  Different
$z^{(a)},z_{a+1},z_{a+2},\cdots$ also lead to different symmetric
polynomials of $\xi_1,\cdots,\xi_a$.  Let $K_{a,m}$ be the number of
linearly independent symmetric polynomials of $\xi_1,\cdots,\xi_a$,
$P_m[z^{(a)},(\xi_1,\cdots,\xi_a);z_{a+1},z_{a+2},\cdots]$, produced
by all the choices of $z^{(a)},z_{a+1},z_{a+2},\cdots$. Let
$F^{(m)}_{\al^{(a)}}(\xi_1,\cdots,\xi_a)$ be a basis of those
symmetric polynomials of $\xi_1,\cdots,\xi_a$, where
$\al^{(a)}=1,2,\cdots,K_{a,m}$ label those linearly independent
symmetric polynomials of $\xi_1,\cdots,\xi_a$.  Thus the different
polynomials of $z^{(a)},z_{a+1},z_{a+2},\cdots$ labeled by
$(\xi_1,\cdots,\xi_a)$ can be written as
\begin{align*}
&\ \ \ P_m[z^{(a)},(\xi_1,\cdots,\xi_a);z_{a+1},z_{a+2},\cdots]
\nonumber\\
&=\sum_{\al^{(a,m)}=1}^{K_{a,m}} F^{(m)}_{\al^{(a,m)}}(\xi_1,\cdots,\xi_a)
P_m[z^{(a)},\al^{(a)};z_{a+1},z_{a+2},\cdots]
\end{align*}
We see that as different polynomials of
$z^{(a)},z_{a+1},z_{a+2},\cdots $ labeled by $\xi_1,\cdots,\xi_a $,
$P_m[z^{(a)},(\xi_1,\cdots,\xi_a);z_{a+1},z_{a+2},\cdots]$ can also
be labeled by $\al^{(a,m)}$.
$F^{(m)}_{\al^{(a,m)}}(\xi_1,\cdots,\xi_a)$ can be viewed as a
conversion factor between the two different labels, $\al^{(a,m)}$
and $\xi_1,\cdots,\xi_a$.

Let $S_a$ be the minimal total power of $z_1,\cdots,z_a$ in $\Phi(\{ z_i \})$.
We see that $K_{a,m}=0$ if $m<S_a$, and
\begin{equation*}
K_a\equiv K_{a,S_a}\neq 0.
\end{equation*}
Since the symmetric polynomial $F^{(m)}_{\al^{(a,m)}}(\xi_1,\cdots,\xi_a)$
of order $m=1$ must have a form
\begin{equation*}
 F^{(m=1)}_{\al^{(a,m)}}(\xi_1,\cdots,\xi_a)\propto \sum \xi_i
\end{equation*}
which vanishes due to the $ \sum \xi_i =0$ condition, we find that
\begin{equation}
\label{Saneq1}
 S_a\neq 1 .
\end{equation}

Here we would like to introduce a unique-fusion condition.  If
$K_{a,m_a}=1$ for all $a$ and allowed $m_a$, then we say that the
symmetric polynomial $\Phi(\{z_i\})$ satisfies the unique-fusion
condition.  In this paper, we will mainly concentrate on symmetric
polynomials $\Phi(\{z_i\})$ that satisfies the unique fusion
condition. In this case we can drop the $\al^{(a)}_i$ labels.

Note that $S_1$ is the minimum power of $z_i$ in $\Phi(\{z_i\})$.  If $S_1>0$,
we can construct another symmetric polynomial with less total power of $z_i$
\begin{equation*}
\t \Phi(\{z_i\})= \Phi(\{z_i\})/\prod_i z_i^{S_1}
\end{equation*}
Since $\prod_i z_i^{S_1}$ is non-zero except at $z=0$,
$\Phi(\{z_i\})$ and $\Phi(\{z_i\})/\prod_i z_i^{S_1}$ have the
same pattern of zeros.  Since the minimal power of $z_i$ in $\t
\Phi(\{z_i\})$ is zero, without loss of generality, we will assume
\begin{equation}
\label{S1eq0}
S_1=0 .
\end{equation}

For any symmetric polynomial, we obtain a sequence of integers
$S_1,S_2,\cdots$.  Different types of symmetric polynomials may
lead to different sequences of integers $S_1,S_2,\cdots$.  Thus we
can use such sequences to characterize different symmetric
polynomials. Since $S_a$ describes how fast a symmetric polynomial
$\Phi(\{z_i\})$ approaches to zero as we bring $a$ variables
together, we will call $S_1,S_2,\cdots$ a pattern of zeros.

For the $\nu=1/q$  Laughlin state \eq{Laughlin}
\begin{equation}
\label{Laughlin}
 \Phi_{1/q} (\{z_i\})=\prod_{i<j} (z_i-z_j)^q ,
\end{equation}
we find that $S_a$ is
\begin{equation}
\label{SaLau}
 S_a=\frac{q a (a-1)}{2}
\end{equation}
For the  Laughlin state, $K_{a,m_a}=0$ or $1$. Thus the  Laughlin
state is a state that satisfies the unique-fusion condition.

\subsection{Derived polynomials}
\label{mdPoly}

In the last section, we have obtained a new polynomial
$P(z^{(a)},\al^{(a)},z_{a+1},\cdots)$ from $\Phi(\{z_i\})$ by fusing $a$
variables together.  We can fuse many clusters of variables together and
obtain, in a similar way, a derived polynomial
\begin{equation*}
 P(\{z_i^{(a)},\al_i^{(a)}\})
\end{equation*}
where $z_i^{(a)}$ is a type-$a$ variable obtained by fusing $a$
$z_i$-variables together.  Since different ways to fuse $z_i$'s
into a $z^{(a)}$ may lead to different derived polynomials, the
index $\al_i^{(a)}$ is introduced to label those different derived
polynomials.

As we let $z_1^{(a)}\to z_1^{(b)}$, we obtain
\begin{align}
\label{fscnd1}
&\ \ \ \
P(z_1^{(a)},\al_1^{(a)};z_1^{(b)},\al_1^{(b)};\cdots)\big|_{z_1^{(a)}\to z_1^{(b)}\equiv z^{(a+b)}}
\\
&\sim (z_1^{(a)}-z_1^{(b)})^{D_{ab}}
\t P(z^{(a+b)},\al^{a+b}; \cdots) + o( (z_1^{(a)}-z_1^{(b)})^{D_{ab}})  .
\nonumber
\end{align}
$D_{ab}$ is another set of data that characterize the symmetric
polynomial $\Phi(\{z_i\})$.  The two sets of data, $S_a$ and
$D_{ab}$ are related by
\begin{equation}
 \label{DabSa}
D_{ab}=S_{a+b}-S_a-S_b.
\end{equation}
This allows us to use $S_a$ to calculate $D_{ab}$.  It also implies that
$D_{ab}$ does not depend on $\al_1^{(a)}$, $\al_1^{(b)}$, and $\al^{(a+b)}$.

To derive \eq{DabSa}, let us write
$\Phi(\{ z_i \})$ as
\begin{align}
\label{PhippP}
&\ \ \ \Phi (z_1,\cdots,z_{a+b};z_{a+b+1},z_{a+b+2},\cdots)
\nonumber\\
& = \la^{S_a}\t\la^{S_b}\sum_{\al^{(a)}, \al^{(b)} }
P(z^{(a)}, \al^{(a)};z^{(b)},\al^{(b)}; z_{a+b+1},\cdots)
\times
\nonumber\\ &\ \ \ \ \ \
F_{\al^{(a)}}(\xi_1,\cdots,\xi_a) F_{\al^{(b)}}(\xi_{a+1},\cdots,\xi_{a+b})
+ \cdots  .
\end{align}
where
\begin{align*}
 z^{(a)}&=(z_1+\cdots+z_a)/a , \ \ \ z^{(b)}=(z_{a+1}+\cdots+z_{a+b})/b  ,
\nonumber\\
\la \xi_i &= z_i-z^{(a)}, \text{ for } i=1,\cdots, a,
\nonumber\\
\t \la \xi_i &= z_i-z^{(b)}, \text{ for } i=a+1,\cdots, a+b.
\end{align*}
In \eq{PhippP}, $\cdots$ represents the higher order terms in
$\la$ and $\t\la$.  Since in $\Phi(\{ z_i \})$, the minimal total
power of $z_1,\cdots,z_{a}$ is $S_a$ and the minimal total power
of $z_{a+1},\cdots,z_{a+b}$ is $S_b$, as a result
$F_{\al^{(a)}}(\xi_1,\cdots,\xi_a)$ and
$F_{\al^{(b)}}(\xi_{a+1},\cdots,\xi_{a+b}) $ are homogenous
polynomial of degree $S_a$ and $S_b$  respectively.  Since the
minimal total power of $z_1,\cdots,z_{a+b}$ is $S_{a+b}$, thus the
minimal total power of $z_1^{(a)}$ and $z_1^{(b)}$ in $P(z^{(a)},
\al^{(a)};z^{(b)},\al^{(b)}; z_{a+b+1},z_{a+b+2},\cdots)$ is
$S_{a+b}-S_a-S_a$.  This proves \eq{DabSa}.

\eq{DabSa} expresses $D_{ab}$ in terms of $S_a$. We can also express $S_a$ in
terms of $D_{ab}$.  Since $S_1=0$, we find that
\begin{equation*}
 S_{a+1}=S_a+D_{a,1}
\end{equation*}
which leads to
\begin{equation}
\label{SaDa1}
 S_a=\sum_{b=1}^{a-1} D_{b,1} .
\end{equation}

\subsection{$n$-cluster condition}

In \Ref{WWsymm}, we have introduced an $n$-cluster condition on
symmetric polynomials $\Phi(\{z_i\})$ to make polynomials of
infinite variables to behave more like polynomials of $n$ variables.
Here we will introduce the $n$-cluster condition in a slightly
different way. A symmetric polynomial $\Phi(\{z_i\})$ satisfies
the $n$-cluster condition
if and only if\\
(a) the fusion of $k n$ variables is unique: $K_{k n,S_{k n}}=1$,
where
$k=$integer,\\
(b) as a function of a $kn$-cluster variable $z^{(k n)}$,
the derived polynomial $P(z^{(k n)}, z^{(a)}_1, \cdots)$
is non-zero if $z^{(k n)}\neq z^{(a)}_i$ (\ie no zeros off the variables
$z^{(a)}_i$).

Let
\begin{equation}
\label{mDn1}
 m\equiv D_{n,1},
\end{equation}
the condition (b) allows us to show that
\begin{equation*}
 D_{k n,a}=k a m .
\end{equation*}
From $D_{n,a}=S_{a+n}-S_a-S_n$,
we find that
$ S_{a+ n} = S_a+  S_n + ma$ which implies that
\begin{equation}
\label{Sakn}
 S_{a+k n} = S_a+ k S_n + mn \frac{k(k-1)}{2}  +kma.
\end{equation}
\eq{Sakn} is the defining relation that defines the $n$-cluster condition on
the pattern of zeros $\{S_a\}$.  We see that $\{S_2,\cdots,S_n\}$ determine
the whole sequence $\{S_a\}$.

\subsection{Occupation description of pattern of zeros}

There are other ways to encode the data $\{S_{a}\}$ which can be
very useful. In the following, we will introduce occupation
description of the pattern of zeros $\{S_{a}\}$.  We first assume
that there are many orbitals labeled by $l=0,1,2,\cdots$. Some
orbitals are occupied by particles and others are not occupied. The
pattern of zeros $\{S_{a}\}$ corresponds to a particular pattern of
occupation. To find the corresponding occupation pattern, we
introduce
\begin{align}
\label{laSa}
l_{a}=S_{a}-S_{a-1},\ \ \ \ \ \ a=1,2,\cdots.
\end{align}
and interpret $l_{a}$ as the label of the orbital that is occupied
by the $a^\text{th}$ particle.  Thus the sequence of integers $l_a$
describe a pattern of occupation. $l_a$ is a monotonically
increasing sequence:
\begin{equation*}
 l_{a+2}-l_{a+1}=S_{a+2}+S_{a}-2S_{a+1} \geq 0
\end{equation*}
as implied by the condition (see \Ref{WWsymm})
\begin{equation}
\label{Del3con}
S_{a+b+c}
- S_{a+b}
- S_{a+c}
- S_{b+c}
+S_{a}
+S_{b}
+S_{c} \geq 0
\end{equation}
with  $b=c=1$.  The $n$-cluster condition \eq{Sakn} becomes
\begin{equation}
\label{lanm}
 l_{a+n}=l_a+m.
\end{equation}
So $\{l_1,\cdots,l_n\}$ determine the whole sequence $\{l_a\}$.
Two sequences, $\{l_1,\cdots,l_n\}$ and $\{S_2,\cdots,S_n\}$, have
a one-to-one correspondence. Thus $\{S_a\}$ and $\{l_a\}$ are
faithful representations of each other.

The occupation distribution can also
be described by another sequence of integers $n_{l}$.  Here $n_{l}$ is the
number of $l_{a}$'s whose value is $l$.  $n_{l}$ is the number of
particles occupying the orbital $l$.  Both $\{l_{a}\}$ and $\{n_{l}\}$ are
equivalent occupation descriptions of the pattern of zeros $\{S_a\}$.

The occupation distribution $\{n_l\}$ (or equivalently $\{l_a\}$)
that describes the pattern of zeros in $\Phi$ has a very physical
meaning.  If we regard $n_l$ as the occupation number on an
orbital described by one-body wave function $\phi_l=z^l$, then the
occupation distribution $\{n_l\}$ will correspond to a many-boson
state described by a symmetric polynomial $\Phi_{\{n_l\}}$.  The
two many-boson states $\Phi$ and $\Phi_{\{n_l\}}$ will have
similar density distribution, in particular in the region far from
$z=0$.

\subsection{Ideal Hamiltonian and zero energy ground state}

For an electron system on a sphere with $N_\phi$ flux quanta, each electron
carries an orbital angular momentum $J=N_\phi/2$ if the electrons are in the
first Landau level.\cite{H8305} For a cluster of $a$ electrons, the maximum
allowed angular momentum is $aJ$. However, for the wave function
$\Phi(\{z_i\})$ described by the pattern of zeros $\{S_a\}$, the maximum
allowed angular momentum for an $a$-electron cluster in $\Phi(\{z_i\})$ is
only $J_a=aJ-S_a$.  The pattern of zeros forbids the appearance of angular
momenta $aJ-S_a+1, aJ-S_a+2,\cdots, aJ$ for any $a$-electron clusters in
$\Phi(\{z_i\})$.

Such a property of the wave function $\Phi(\{z_i\})$ can help us to construct
an ideal Hamiltonian such that $\Phi$ is the exact zero-energy ground
state.\cite{H8305,GWW9267,W9355,RR9984,R0634,SRC0760} The ideal Hamiltonian
has the following form
\begin{equation}
\label{Hideal}
 H_{\{S_a\}}=\sum_a \sum_{a\text{-electron clusters}}
(P^{(a)}_{S_a} + P_{\bar\cH_a}) ,
\end{equation}
where $P^{(a)}_S$ is a projection operator that projects onto the subspace of
$a$ electrons with total angular momenta $aJ-S+1,\cdots, aJ$ and $P_{\bar\cH_a}$
is a projection operator into the space $\bar\cH_a$.

What is the space $\bar\cH_a$? Let us fix $z_{a+1},\ z_{a+2},\
\cdots$, and view the wave function $\Phi(z_1,\cdots,z_N)$ as a
function of $z_1,\cdots,z_a$. Such a wave function describes an
$a$-electron state.  If $\Phi(z_1,\cdots,z_N)$ is described by a
pattern of zeros $\{S_a\}$, the $a$-electron state defined above
can have a non zero projection into the space $\cH_{a,S_a}$, where
$\cH_{a,S_a}$ is a space with a total angular mentum $aJ-S_a$.
However, different positions of other electrons may lead to
different projections of the $a$-electron states in the space
$\cH_{a,S_a}$. $\cH_a$ is then the subspace of $\cH_{a,S_a}$ that
is spanned by those states. $\bar\cH_a$ is the subspace of
$\cH_{a,S_a}$ formed by vectors that are orthogonal to $\cH_a$.

We see that by construction, $\Phi$ described by a pattern of zeros $\{S_a\}$
is a zero energy ground state of the ideal Hamiltonian $H_{\{S_a\}}$.
However, the zero energy states of $H_{\{S_a\}}$ are not unique.  In fact, any
state $\t \Phi$ with a pattern of zeros $\{\t S_a\}$ will be a zero-energy
eigenstate of $H_{\{S_a\}}$ if $\t S_a\geq S_a$.  (If $\t S_a=S_a$, we will
further require that $H_a \subseteq \t H_a$.)

But on sphere, $H_{\{S_a\}}$ may have a unique ground state.
If the number of electrons is a multiple of $n$: $N=nN_c$, and the number of
the flux quanta $N_\phi$ satisfies
\begin{align}
\label{NphiN}
 2J=N_\phi &= \frac{2S_n}{n}+m(N_c-1) ,
\end{align}
then the state $\Phi$ can be put on a sphere and corresponds to a unique
uniform state with zero total angular momentum.  Such a state may be the
unique zero-energy ground state of the ideal Hamiltonian $H_{\{S_a\}}$ on the
sphere.

If we increase $N_\phi$ (but fix the number of particles $N$), then
$H_{\{S_a\}}$ will have more zero-energy states.  Those zero-energy
states can be viewed as formed by a few quasiparticle excitations.
Those quasiparticle excitations may carry fraction charges and
fractional statistics.  In the next section, we will study those
zero-energy quasiparticles through the pattern of zeros.

\section{Quasiparticles}
\label{qp}

\subsection{Quasiparticles and patterns of zeros}
\label{qPOZ}

If we let $z_1,\cdots,z_a$ approach $0$ in a ground state wave
function $\Phi(\{z_i\})$, we obtain
\begin{align*}
 \Phi(\{z_i\})|_{\la\to 0}=\la^{S_a} P(\xi_1,\cdots,\xi_a;z_{a+1},\cdots)+\cdots
\end{align*}
where $z_i=\la\xi_i$, $i=1,\cdots,a$.  The pattern of zeros
$\{S_a\}$ indicates that the $z=0$ point is the same as any other
point, since we will get the same pattern of zeros if we let
$z_1,\cdots,z_a$ to approach to any other point.  Thus the pattern
of zeros $\{S_a\}$ describes a state with no quasiparticle at
$z=0$.

If a new symmetric polynomial $\Phi_\ga(\{z_i\})$ has a quasiparticle at
$z=0$, $\Phi_\ga(\{z_i\})$ will have a different pattern of zeros
$\{S_{\ga;a}\}$ as $z_1,\cdots,z_a$ approach $0$:
\begin{align}
\label{Sgaa}
\Phi_\ga(\{z_i\})=\la^{S_{\ga;a}}
P_\ga(\xi_1,\cdots,\xi_a;z_{a+1},\cdots)+\cdots  .
\end{align}
In other words, the minimal total powers of $z_1,\cdots,z_a$ in
$\Phi_\ga(\{z_i\})$ is $S_{\ga;a}$.  Thus we can use $\{S_{\ga;a}\}$ to
characterize different quasiparticles.  Here the index $\ga$ labels different
type of quasiparticles. We would like to stress that the quasiparticle wave
function $\Phi_\ga(\{z_i\})$ is still described by the pattern of zeros
$\{S_a\}$ if we let $z_1,\cdots,z_a$ approach to any other point away from
$z=0$.

Just as $\{S_a\}$, $\{S_{\ga;a}\}$ should also satisfy certain conditions.
Here, we will consider quasiparticle states $\Phi_\ga(\{z_i\})$ that have a
zero energy for the ideal Hamiltonian $H_{\{S_a\}}$.
The zero-energy condition on quasiparticles
requires that $S_{\ga;a}$ should satisfy
\begin{equation}
\label{SqpS}
 S_{\ga;a}\geq S_{a} .
\end{equation}
Although both the ground state $\Phi$ and the quasiparticle states $\Phi_\ga$
have a zero energy, the ground state has the minimal power of $z_i$ while the
quasiparticle states have higher total powers of $z_i$.

\begin{figure}[t]
\centerline{
\includegraphics[scale=0.5]{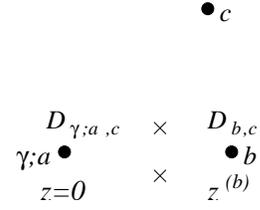}
}
\caption{
As a function of $z^{(c)}$,
$P'_\ga(\al^{(a)};z^{(b)},\al^{(b)};z^{(c)},\al^{(c)};\cdots)$ has  a
$D_{\ga;a,c}$th order zero as $z^{(c)}\to 0$, and a $D_{b,c}$th order zero as
$z^{(c)}\to z^{(b)}$.  The crosses mark the positions of other zeros that are
not at $z=0$ and not at $z^{(b)}$.
}
\label{qpabc}
\end{figure}

In the following, we will discuss other conditions on
$S_{\ga;a}$.  Let $P_\ga(\{z^{(a)}_i,\al^{(a)}_i\})$ be the derived polynomial
obtained from $\Phi_\ga(\{z_i\})$. First we let $z^{(a)}\to 0$ in $P_\ga$:
\begin{align*}
&\ \ \ \ P_\ga(z^{(a)},\al^{(a)};z^{(b)},\al^{(b)};\cdots)
\nonumber\\
&\sim (z^{(a)})^{D_{\ga;a}} P'_\ga(\al^{(a)};z^{(b)},\al^{(b)};\cdots) .
\end{align*}
Then we let $z^{(b)}\to 0$
and let $D_{\ga;a,b}$
be the order of zeros in
$P'_\ga(\al^{(a)};z^{(b)},\al^{(b)};\cdots)$ as
$z^{(b)}\to 0$:
\begin{align*}
P'_\ga(\al^{(a)};z^{(b)},\al^{(b)};\cdots)
\sim (z^{(b)})^{D_{\ga;a,b}} P''_\ga(\al^{(a)};\al^{(b)};\cdots) .
\end{align*}
We have
\begin{equation}
\label{DgaSS}
S_{\ga;a+b}= S_{\ga;a} + D_{\ga;a,b} +S_b ,
\end{equation}
which leads to a condition on $S_{\ga;a}$
\begin{equation*}
 S_{\ga;a+b} - S_{\ga;a} - S_b \geq 0 .
\end{equation*}

To get more conditions on $S_{\ga;a}$, let us consider fusing three variables
$z^{(a)}$, $z^{(b)}$, and $z^{(c)}$ to $0$.  We may first fuse $z^{(a)}$ to 0
to obtain $P'_\ga(\al^{(a)};z^{(b)},\al^{(b)};z^{(c)},\al^{(c)};\cdots) $.
Then we fuse $z^{(b)}$ to 0 to obtain
$P''_\ga(\al^{(a)};\al^{(b)};z^{(c)},\al^{(c)};\cdots) $.  Last we fuse
$z^{(c)}$ to 0.  In this case, we obtain a $D_{\ga;a+b,c}$th order zero as
$z^{(c)}\to 0$ in $P''_\ga(\al^{(a)};\al^{(b)};z^{(c)},\al^{(c)};\cdots) $. We
also know that $P'_\ga$ has a $D_{b,c}$th order zero as $z^{(c)}\to z^{(b)}$
and a $D_{\ga;a,c}$th order zero as $z^{(c)}\to 0$ (see \ref{qpabc}). If
$z^{(b)}$ is very close to 0, we find that $D_{\ga;a+b,c}$ is the sum of
$D_{\ga;a,c}$, $D_{b,c}$, and the zeros close to $z=0$ but not at $z=0$ and at
$z^{(b)}$.  This way, we find
\begin{align}
\label{del3qp}
 D_{\ga;a+b,c}\geq D_{\ga;a,c}+D_{b,c}
\end{align}
which implies that
\begin{align}
\label{Del3qp}
S_{\ga;a+b+c}
- S_{\ga;a+b}
- S_{\ga;a+c}
- S_{b+c}
+S_{\ga;a}
+S_{b}
+S_{c} \geq 0 .
\end{align}
\eq{Del3qp} generalizes the condition \eq{Del3con}
on the pattern of zeros of the ground state.

The $n$-cluster condition implies that, as a function of $z^{(n)}$,
$P_\ga(z^{(n)}, z_i^{(a)},\cdots)$ is non-zero if $z^{(n)}\neq z^{(a)}_i$ and
$z^{(n)}\neq 0$.  Therefore, the inequality \eq{del3qp} is saturated when
$c=n$:
\begin{align*}
 D_{\ga;a+b,n}= D_{\ga;a,n}+D_{b,n} ,
\end{align*}
which implies that (setting $a=0$)
\begin{align*}
 S_{\ga;b+n} &= S_{\ga;b} + D_{\ga;0,n} + S_{b+n}-S_b
\nonumber\\
 &=S_{\ga;b} + S_n+D_{\ga;0,n} + m b
,
\end{align*}
or
\begin{align*}
S_{\ga;a+kn}
=S_{\ga;a} + k(S_n + ma + D_{\ga;0,n})+mn\frac{k(k-1)}{2} .
\end{align*}
Since $S_\ga=0$, we have $S_{\ga;n}=S_n+D_{\ga;0,n}$. Thus
\begin{align}
\label{Sqan}
S_{\ga;a+kn}
=S_{\ga;a} + k(S_{\ga;n} + ma )+mn\frac{k(k-1)}{2} .
\end{align}
for $a\geq 0$. This is the $n$-cluster condition on $S_{\ga;a}$.

\subsection{Solutions of patterns of zeros for quasiparticles}
\label{qppoz}

The ground state of the ideal Hamiltonian \eq{Hideal} is a state
described by a pattern of zero $\{S_a\}$.  The zero-energy
quasiparticles above the ground state $\Phi$ can be characterized
by a pattern of zeros $\{S_{\ga;a}\}$.  If $\Phi$ is a state with
the $n$-cluster form, then all the $S_{\ga;a}$'s are determined
from $S_{\ga;1},\cdots,S_{\ga;n}$ (see \eq{Sqan}).  Thus
zero-energy quasiparticles are labeled by $n$ integers:
$S_{\ga;1},\cdots,S_{\ga;n}$.  From the discussion in the last
section, we find that those integers should also satisfy
\begin{align}
\label{SqCond}
& S_{\ga;a+b}- S_{\ga;a} - S_b \geq 0,
\\
&S_{\ga;a+b+c}
- S_{\ga;a+b}
- S_{\ga;a+c}
- S_{b+c}
+S_{\ga;a}
+S_{b}
+S_{c} \geq 0 .
\nonumber
\end{align}
Numerical experiments suggest that the solutions of \eq{SqCond} always satisfy
\eq{SqpS}. Thus the condition \eq{SqpS} is not included.

For a given pattern of zeros $\{S_a\}$ describing a symmetric
polynomial $\Phi$ of $n$-cluster from, \eq{SqCond} has many
solutions that satisfy the $n$-cluster condition \eq{Sqan}. Those
solutions can be grouped into equivalent classes.

To describe such equivalent classes, we need to use
the occupation description of the pattern of zeros $\{S_{\ga;a}\}$.
We introduce
\begin{align*}
l_{\ga;a}=S_{\ga;a}-S_{\ga;a-1},\ \ \ \ \ \ a=1,2,\cdots.
\end{align*}
and interpreted $l_{\ga;a}$ as the label of the orbital that is
occupied by the $a^\text{th}$ particle.  Thus the sequence of
integers $l_{\ga;a}$ describes a pattern of occupation.  We note
that $l_{\ga;a+1}=D_{\ga;a,1}$. \eq{del3qp} implies that
$l_{\ga;a+1} \geq l_{\ga;a}$.  The occupation distribution can
also be described by another sequence of integers $n_{\ga;l}$.
Here $n_{\ga;l}$ is the number of $l_{\ga;a}$'s whose value is
$l$. $n_{\ga;l}$ is the number of particles occupying the orbital
$l$. Both $\{l_{\ga;a}\}$ and $\{n_{\ga;l}\}$ are equivalent
occupation descriptions of the pattern of zeros $\{S_{\ga;a}\}$.

The $n$-cluster condition on $\{S_{\ga;a}\}$ implies that
\begin{equation}
\label{lgancl}
 l_{\ga;a+n}= l_{\ga;a}+m ,
\end{equation}
for $a\geq 1$. This implies that occupation numbers $\{n_{\ga;l}\}$ satisfy
\begin{equation}
\label{ngancl}
 n_{\ga;l+m}=n_{\ga;l}
\end{equation}
for a large enough $l$.  Thus the occupation numbers $n_{\ga;l}$ become a
periodic function of $l$ with a period $m$ in large $l$ limit.
In that limit, each $m$ orbitals contain $n$ particles.

Now consider two quasiparticles $\ga_1$ and $\ga_2$ described by two sequences
of occupation numbers $\{n^{\ga_1}_l\}$ and $\{n^{\ga_2}_l\}$.  If
$n_{\ga_1;l}=n_{\ga_2;l}$ for large enough $l$, then the density distributions
of two many-boson states $\Phi_{\ga_1}$ and $\Phi_{\ga_2}$ only differ by an
integral number of electrons near $z=0$.  Thus the two quasiparticles $\ga_1$
and $\ga_2$ differ only by an integral number of electrons.  This motivates us
to say that the two quasiparticles to be equivalent.

Numerical experiments suggest that each equivalent class
is represented by a simple occupation distribution $n_{\ga;l}$ that
is periodic for all $l>0$:
\begin{equation}
\label{ngalcan}
 n_{\ga;l+n}=n_{\ga;l},\ \ \ \ \  l\geq 0.
\end{equation}
We call such a distribution the canonical distribution for the corresponding
equivalent class.
For a canonical occupation distribution, there are $n$ particles in every unit
cell $(l=0,\cdots, m-1)$, $(l=m,\cdots, 2m-1)$, $\cdots$.

We can obtain all the equivalent classes of the  quasiparticles by finding all
the canonical distributions that satisfy \eq{SqCond}.  The equivalent classes
of the quasiparticles correspond to fractionalized excitations.

\section{Topological properties from pattern of zeros}

A FQH state characterized by a pattern of zeros $\{S_a\}$ can have many
topological properties, such as quasiparticle quantum
numbers,\cite{L8395,ASW8422} ground state degeneracy on compact
space,\cite{HR8529,Wtop,WNtop} and edge excitations.\cite{Wedgerev}  In this
section, we are going to calculate some of those topological properties from
the data $\{S_a\}$.

\subsection{Charge of quasiparticles}

We have seen that a quasiparticle excitation labeled by $\ga$ are
characterized by a sequence of integers: $\{S_{\ga;a}\}$.  We would
like to calculate the quantum numbers of the quasiparticle from
$\{S_{\ga;a}\}$.

To calculate the quasiparticle charge, we compare the occupations $n_l$ that
describe the pattern of zeros $\{S_a\}$ of the ground state $\Phi$ and the
occupations $n_{\ga;l}$ that describe the pattern of zeros $\{S_{\ga;a}\}$ of
a quasiparticle state $\Phi_\ga$.  We divide $l=0,1,2,\cdots$ into unit cells
each containing $m$ orbitals: $l=0,\cdots,m-1;m,\cdots,2m-1;\cdots$.  $n_l$
and $n_{\ga;l}$ contain the same number of particles in the $k$th unit cell if
$k$ is large enough.  Since $n_l$ is a distribution that corresponds to zero
quasiparticle charge, we might think that the quasiparticle charge
corresponding to the distribution $n_{\ga;l}$ is given by
\begin{equation*}
 \sum_{l=0}^{km} (n_l-n_{\ga;l})
\end{equation*}
in large $k$ limit.  However, this result is incorrect. Although
$n_l$ and $n_{\ga;l}$ contain the same $n$ particles in the $k$th
unit cell (for a large $k$), the ``centers of the mass'' of the
two distributions in the $k$th cell are different.  The shift of
the ``centers of mass'' is given by
\begin{equation*}
\frac 1 n \sum_{l=km-m}^{km-1} (n_l-n_{\ga;l}) l
\end{equation*}
Shifting the ``center of mass'' by $m$ is equivalent to
adding/removing $n$ particles. Thus the total quasiparticle charge
is given by
\begin{align}
\label{Qqnlq}
Q_\ga=
\sum_{l=0}^{km} (n_l-n_{\ga;l})-\frac 1 m \sum_{l=km-m}^{km-1} (n_l-n_{\ga;l}) l,
\end{align}
for a large enough $k$.  Note that in the above definition, a
charge $+1$ corresponds to an absence of an electron.  For a
canonical occupation distribution satisfying \eq{ngalcan}, the
first term $\sum_{l=0}^{km} (n_l-n_{\ga;l})$ vanishes.

Since the two descriptions of occupation distributions,
$\{l_{\ga;a}\}$ and $\{n_{\ga;l}\}$, have a one-to-one
correspondence, we can also express $Q_\ga$ in terms of
$\{l_{\ga;1},\cdots,l_{\ga;n}\}$ (note that, according to
\eq{lgancl}, $\{l_{\ga;1},\cdots,l_{\ga;n}\}$ determines the whole
sequence $\{l_{\ga;a}\}$):
\begin{equation}
\label{Qgala}
 Q_\ga= \frac{1}{m} \sum_{a=1}^{n} (l_{\ga;a}-l_a) .
\end{equation}

There is another way to calculate the charge of a quasiparticle.  We can put
the quasiparticle state $\Phi_\ga$ on a sphere with $N_\phi$ flux quanta. If
we move the quasiparticle around a loop that spans a solid angle $\Om$, the
quasiparticle state $\Phi_\ga$ will generate a Berry's phase
\begin{equation}
\label{thBQq}
 \th_B (\Om) = \frac{N_\phi \Om}{2} Q_\ga + O(N_\phi^0)
\end{equation}
The part in $\th_B$ that is proportional to $N_\phi$ allows us to
determine the charge $Q_\ga$.

\begin{figure}[t]
\centerline{
\includegraphics[scale=0.6]{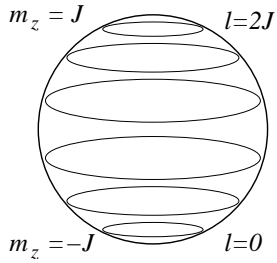}
}
\caption{
The $2J+1$ orbital on a sphere form a angular momentum $J$ representation of
the $O(3)$ rotation.
}
\label{mzl}
\end{figure}

From the Berry's phase $\th_B (\Om) $ (as a function of the solid angle $\Om$),
we can find out the angular momentum $J_q$ of the quasiparticle
\begin{equation*}
 J_q=\th_B(\Om)/\Om ,
\end{equation*}
(note that $\th_B\propto \Om$.) The occupation distribution $n_l$
for the ground state $\Phi$ describes a trivial quasiparticle with
zero charge and zero angular momentum.  The occupation
distribution $n_{\ga;l}$ for the quasiparticle state $\Phi_\ga$
describes a non-trivial quasiparticle with a non-zero angular
momentum $J_q$. Since the orbital $\phi_l$, when put on a sphere
with $N_\phi$ flux quanta, is identified with an angular momentum
eigenstate $(J,J_z)=(\frac{N_\phi}{2}, l-J)$, the quasiparticle
state $\Phi_\ga$ is an eigenstate of total $J^\text{tot}_z$ (see
Fig. \ref{mzl}). Since $\Phi_\ga$ is the state with maximal
$J^\text{tot}_z$, the total angular momentum of the quasiparticle
is $J^\text{tot}=J^\text{tot}_z\equiv J_q$.

Since the occupation distribution $n_l$ describes a state with
$J^\text{tot}_z=0$, the angular momentum of the quasiparticle is
\begin{equation*}
 J_q=\sum_{l=0} (n_{\ga;l}-n_l) (l-J) ,
\end{equation*}
or more generally the Berry's phase of the quasiparticle is
\begin{equation}
\label{thBq}
\th^\ga_B=\Om \sum_{l=0} (n_{\ga;l}-n_l) (l-J) .
\end{equation}
where the upper bound of the summation $\sum_{l=0}$ is roughly at
$l=N_\phi=2J$.  From the part of $\th^\ga_B$ that is linear in
$N_\phi=2J$, we can recover \eq{Qqnlq} for the charge of the
quasiparticle.

\subsection{Orbital spin of quasiparticles}

We have seen that the Berry's phase of the quasiparticle contains
a term linear in $N_\phi=2J$ that is related to the quasiparticle
charge. The Berry's phase also contains a constant term which by
definition determines the orbital spin $S_\ga^\text{osp}$ of the
quasiparticle.\cite{WZspv,WZspvE,Wtoprev} More precisely, in the
large $N_\phi=2J$ limit, we have
\begin{align}
\label{thBSosp}
 \frac{\th^\ga_B}{\Om}
= J Q_\ga + S_\ga^\text{osp} + O(J^{-1}) .
\end{align}
Therefore, to calculate $S_\ga^\text{osp}$, we need to evaluate
\eq{thBq} carefully. \eq{thBq} is not a well defined expression
since the upper bound of the summation $\sum_{l=0}$ is not given
precisely (on purpose).  So to evaluate \eq{thBq}, we first
regulate \eq{thBq} as
\begin{equation}
\label{thBqR}
\th^\ga_B=\Om \sum_{l=0} (n_{\ga;l}-n_l) (l-J) \e^{-\al^2 l^2} ,
\end{equation}
and then take the small $\al$ and large $J$ limit with $\al\sim
1/J$. The key difference between \eq{thBq} and \eq{thBqR} is that
\eq{thBqR} has a soft (or smooth) cut-off near the upper bound of
the summation $\sum_{l=0}$. After evaluating the regulated
summation \eq{thBqR} in appendix \ref{calosp}, we find that the
orbital spin of the quasiparticle is given by
\begin{equation}
\label{ospnl}
 S_\ga^\text{osp}=\sum_{l=0}^{m-1} (n_{\ga;l} - n_l)
\Big(\frac{l}{2}-\frac{l^2}{2m}\Big)  .
\end{equation}

\eq{ospnl} is valid only for canonical occupation distributions that satisfy
\eq{ngalcan}. In terms of $l_{\ga;a}$, we can rewrite \eq{ospnl} as
\begin{equation}
\label{ospla}
 S_\ga^\text{osp}=\sum_{a=1}^n \frac{ml_{\ga;a}-l_{\ga;a}^2-ml_a+l_a^2}{2m} .
\end{equation}
However, \eq{ospla} is more general. It applies to generic
quasiparticles, even those whose occupation distributions do not
satisfy \eq{ngalcan}.

We would like to stress that \eqn{ospla} is
obtained through an untested method. Its validity is confirmed
only for Abelian quasiparticles (see appendix \ref{ospAb}).
Independent confirmation is needed for more general cases.

\subsection{Ground state degeneracy on torus}

A FQH state has a topological ground state degeneracy on torus,
which is robust against any local perturbations.\cite{Wtop, WNtop}
Such a topological ground state degeneracy is part of the defining
properties of topological orders.

According to topological quantum field theory,\cite{W8951,RSW,DFL} the
topological ground state degeneracy is equal to the number of
quasiparticle types.  To be precise, two quasiparticles are
regarded equivalent if they differ by a multiple of electrons.
Thus, the topological ground state degeneracy is equal to the
number of equivalent classes of quasiparticles introduced in
section \ref{qppoz}.

To understand such a result, let us consider a quasiparticle state $\Phi_\ga$
on a sphere (which is a zero-energy state of the ideal Hamiltonian).  We
stretch the sphere into a thin long tube.  The state $\Phi_\ga$ remains to be
a zero-energy state in such a limit.  According to
\Ref{SL0604,BKW0608,SY0802}, the FQH state $\Phi_\ga$ in such a limit becomes
a charge-density-wave (CDW) state characterized by a particle occupation
distribution among the orbitals on the thin tube.  We expect such an
occupation distribution is given by $n_{\ga;l}$ in the large $l$ limit.  We
see that $n_{\ga;l}$ in the large $l$ limit is a CDW state in the thin
cylinder limit that is compatible with the zero-energy requirement.  This
suggests that the quasiparticle types (determined by $n_{\ga;l}$ in the large
$l$ limit) and the zero-energy CDW states on thin cylinder (given by
$n_{\ga;l}$ in the large $l$ limit) are closely related.

The above result allows us to construct zero energy ground states on
a torus. Let us consider a torus with $N_\phi=N_c m$ flux quanta.
There are $N_\phi$ orbitals on such a torus. Those orbitals are
labeled by $l=0,1,2,\cdots,N_\phi-1$. Now let us consider a $N=N_c
n$-electron FQH state on such a torus.  The FQH state is described
by a pattern of zeros $\{S_a\}$ of $n$-cluster form and has a
filling fraction $\nu=n/m$.  What are the degenerate ground states
of such a FQH state on the torus?

According to \Ref{SL0604,BKW0608,SY0802}, the zero-energy ground
state wave functions of a FQH state in the thin cylinder limit can
be described by certain occupation distribution patterns (or certain
CDW states). The above discussion on sphere suggests that the
canonical distributions $n_{\ga;l}$ for the quasiparticles are just
those occupation distributions. (Note that the canonical
distributions fill each $m$ orbitals with $n$ electrons and give
rise to very uniform distributions.) Thus each canonical
distribution $n_{\ga;l}$ gives rise to a $N$-electron ground state
wave function on the thin torus which corresponds to a degenerate
ground state.  We see that the canonical occupation distributions
$n_{\ga;l}$ characterize both the degenerate ground states on torus
and different types of quasiparticles. This explains why the ground
state degeneracy on torus is equal to quasiparticle types. We also
see that we can use $m$ integers $n_{\ga;l}$, $l=0,1,\cdots,m-1$, to
label different degenerate ground states.

The condition on the CDW distributions, $n_{\ga;l}$, that correspond to the
zero energy ground states on thin torus can be stated in the translation
invariant way.  Using $l_{\ga;a}$'s, we can rewrite the second expression of
\eqn{SqCond} as
\begin{align}
\label{laCond}
 \sum_{k=1}^c (l_{\ga;a+b+k} - l_{\ga;a+k}) &\geq S_{b+c}-S_b-S_c=D_{b,c}  .
\end{align}
To understand the meaning of \eqn{laCond}, let us set $c=1$ in the
in \eqn{laCond}.  In this case, \eqn{laCond} requires that a zero
energy CDW distribution must satisfy the
following condition: any groups of $b$ electrons must spread over $D_{b,1}+1$
orbitals or more.

After knowing the one-to-one correspondence between the
quasiparticle types and the degenerate ground states, we like to ask
which quasiparticle type correspond to which ground states? Since
both  quasiparticle types and the degenerate ground states are
labeled by the canonical occupation distribution, one may expect
that a quasiparticle labeled by $n_{\ga;l}$ will correspond to the
ground state labeled by $n_{\ga;l}$.  However, this does not have to
the case. In general, a quasiparticle labeled by $n_{\ga;l}$ may
correspond to the ground state labeled by $n_{\ga;l+l_s}$.  Later,
we see that the shift $l_s$ is indeed non-zero. So we will denote
the ground state wave function that corresponds to a quasiparticle
$\ga$ as $\Phi_{\{n_{\ga;l+l_s}\}}(\{\v X_i\})$ or more briefly as
$\Phi_\ga(\{\v X_i\})$.

\subsection{Quantum numbers of ground states}

The Hamiltonian of FQH state on torus has certain symmetries. The
degenerate ground state on torus will form a representation of
those symmetries. In this section, we will discuss some of those
representations.

\subsubsection{The Hamiltonian on torus}

First, let us specify the Hamiltonian of the FQH system more carefully.
The kinetic energy of the FQH system is determined by the
following one-electron Hamiltonian
on a torus $(X^1,X^2)\sim (X^1+1,X^2)\sim (X^1,X^2+1)$ with a general mass matrix:
\begin{equation}
\label{HAXAY}
   H_{X} = - \frac{1}{2}\sum_{i,j=1,2}
\Big(\frac{\partial}{\partial X^i} - \imth A_i \Big)
g_{ij}
\Big(\frac{\partial}{\partial X^j} - \imth A_j \Big) ,
\end{equation}
where $g$ is the inverse-mass-matrix:
\begin{equation}
\label{mtau}
 g(\tau)=
\bpm
\tau_y+\frac{\tau_x^2}{\tau_y} & -\frac{\tau_x}{\tau_y}\\
-\frac{\tau_x}{\tau_y} & \frac{1}{\tau_y}
\epm
\end{equation}
and
\begin{equation}
\label{AXAY}
(A_1,A_2)=(-2\pi N_\phi X^2, 0)
\end{equation}
gives rise to a uniform magnetic field
with $N_\phi$ flux quanta going through the torus.
The state in the first Landau level has a form
\begin{align}
\label{PsiX}
\phi(X^1,X^2) &= f(X^1+\tau X^2) \e^{\imth\pi N_\phi \tau (X^2)^2}
\end{align}
where $\tau=\tau_x+\imth\tau_y$ and $f(z)$ is a holomorphic function that
satisfies the following periodic boundary condition
\begin{equation}
\label{qpcon}
  f(z+a+b\tau)=f(z) \e^{-\imth\tau \pi b^2 N_\phi-\imth 2\pi b N_\phi z}
,\ \ \ \ \
a,b=\text{ integers} .
\end{equation}
The above holomorphic functions can be
expanded by the following $N_\phi$ basis wave functions:
\begin{equation}
\label{thNphi}
 f^{(l)}(z|\tau)
=\sum_{k} \e^{ \imth\frac{\pi \tau}{N_\phi} (N_\phi k + l)^2
+ \imth 2  \pi (N_\phi k +l) z },
\end{equation}
where $l=0,\cdots,N_\phi-1$.
The corresponding wave functions
\begin{equation}
\label{philX}
\phi^{(l)}(X^1,X^2|\tau)=
f^{(l)}(X^1+\tau X^2|\tau) \e^{\imth \pi N_\phi \tau (X^2)^2}
\end{equation}
are orbitals on the torus.

\begin{figure}[t]
\centerline{
\includegraphics[scale=0.55]{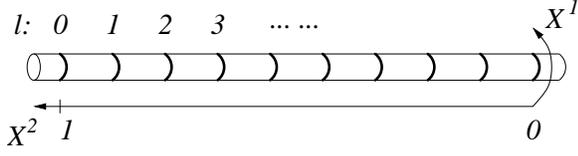}
} \caption{ The circular orbital wave function $\phi^{(l)}(X^1,X^2)$
in the thin cylinder limit. A think line marks the positions where
$\phi^{(l)}(X^1,X^2)$ is peaked. } \label{thncyl}
\end{figure}

The thin cylinder limit\cite{SL0604,BKW0608,SY0802,ABK0875}
is realized by taking $\tau_y\to \infty$.
In such a limit
\begin{align*}
\phi^{(l)}(X^1,X^2|\tau) &=
f^{(l)}(X^1+\tau X^2|\tau) \e^{\imth \pi N_\phi \tau (X^2)^2}
\nonumber\\
&=\sum_{k}\e^{\imth\pi\tau N_\phi(k+\frac{l}{N_\phi}+X^2)^2
+ \imth 2  \pi (N_\phi k +l) X^1 }
\nonumber\\
&\approx
\e^{\imth\pi\tau N_\phi[X^2-(1-\frac{l}{N_\phi})]^2
+ \imth 2  \pi (l-N_\phi) X^1 }
.
\end{align*}
We see that in the thin cylinder limit, the orbital wave function
$\phi^{(l)}(X^1,X^2|\tau)$ is non-zero only near $X^2=1-\frac{l}{N_\phi}$ (see
Fig.  \ref{thncyl}).

\subsubsection{Translation symmetry}

First let us consider the symmetry of the Hamiltonian \eq{HAXAY}.
Since the magnetic field is uniform, we expect the translation symmetry
in both $X^1$- and $X^2$-directions.
The Hamiltonian \eq{HAXAY} does not depend on $X^1$, thus
\begin{equation*}
 T^\dag_{(d_1,0)} H T_{(d_1,0)} =H,\ \ \ \ \ \
T_{(d_1,0)}=\e^{d_1 \prt_{X^1}}
\end{equation*}
But the Hamiltonian \eq{HAXAY} depends on $X^2$ and there seems no
translation symmetry in $X^2$-direction.  However, we do have a
translation symmetry in $X^2$-direction once we include the gauge
transformation. The Hamiltonian \eq{HAXAY} is invariant under
$X^2\to X^2+d_2$ transformation \emph{followed} by a $\e^{\imth
2\pi N_\phi d_2 X^1+\imth \phi}$ $U(1)$-gauge transformation:
\begin{equation*}
T^\dag_{(0,d_2)} H T_{(0,d_2)} =H,\ \ \ \ \ \
T_{(0,d_2)}=\e^{\imth 2\pi N_\phi  d_2 X^1+\imth \phi}\e^{d_2 \prt_{X^2}} .
\end{equation*}
In general
\begin{equation}
\label{TdH}
T^\dag_{\v d} H T_{\v d} =H,\ \ \ \ \ \
T_{\v d}=\e^{\imth 2\pi N_\phi d_2 X^1+\imth \pi N_\phi d_1d_2}\e^{\v d \cdot \v \prt} .
\end{equation}
where we have chosen the constant phase in $T_{\v d}$
\begin{equation}
\label{phiBdxdy}
 \phi=\pi N_\phi d_xd_y
\end{equation}
to simplify the later calculations.
The operator $T_{\v d}$ is called magnetic translation operator.  So the
Hamiltonian \eq{HAXAY} does have translation symmetry in any directions. But
the (magnetic) translations in different directions do not commute
\begin{equation}
\label{TdTd}
 T_{\v d} T_{\v d'}
=\e^{\imth 2\pi N_\phi (d_1d'_{2}-d_2d'_1)} T_{\v d'} T_{\v d}  .
\end{equation}
and momenta in $X^1$- and $X^2$-directions cannot be well defined at the same
time.

However, when $N_\phi$ is an integer, $T_{(1,0)}$ and $T_{(0,1)}$
do commute and the wave function $\phi^{(l)}$  in the first Landau
levels satisfies
\begin{equation*}
 T_{(1,0)} \phi^{(l)}= T_{(0,1)} \phi^{(l)}= \phi^{(l)} .
\end{equation*}
Therefore $\phi^{(l)}$ is a wave function that lives on the torus
$(X^1,X^2)\sim (X^1+1,X^2)\sim (X^1,X^2+1)$.

On a torus, the allowed translations are
discrete since those translation must commute with $T_{(1,0)}$ and $T_{(0,1)}$.
The smallest translation in $X^1$ and $X^2$ directions are given by
\begin{equation*}
 T_1=T_{(\frac{1}{N_\phi},0)},\ \ \ \ \ \
 T_2=T_{(0,\frac{1}{N_\phi})},
\end{equation*}
which satisfies
\begin{equation}
\label{magtrans}
 T_1T_2=\e^{\imth 2\pi/N_\phi} T_{2} T_{1} .
\end{equation}
We also find that (see appendix \ref{T1T2act})
\begin{equation*}
 T_1\phi^{(l)}=\e^{\imth 2\pi l/N_\phi} \phi^{(l)},\ \ \ \ \
 T_2\phi^{(l)}= \phi^{(l+1)}  .
\end{equation*}

The many-body ground state wave functions
$\Phi_{\{n_{\ga;l+l_s}\}}(\{\v X_i\}) = \Phi_\ga(\{\v X_i\})$. of
the FQH state on torus are labeled by the canonical occupation
distributions (see \eq{ngalcan}). In the thin cylinder limit
$\tau_y\to \infty$,
the many-body ground state wave functions
$\Phi_{\{n_{\ga;l+l_s}\}}$ become the CDW wave functions described
by the occupation distributions $\{n_{\ga;l+l_s}\}$ where there
are $n_{\ga;l+l_s}$ electrons occupying the orbital $\phi^{(l)}$.
This allows us to obtain how $\Phi_{\{n_{\ga;l+l_s}\}}$'s
transform under translation:
\begin{align*}
 T_1\Phi_{\{n_{\ga;l+l_s}\}}&=
\e^{\imth 2\pi \sum_{l=0}^{N_\phi-1} n_{\ga;l+l_s} l/N_\phi}
\Phi_{\{n_{\ga;l+l_s}\}}
\nonumber\\
&=\e^{\imth 2\pi \sum_{l=0}^{N_\phi-1} n_{\ga;l} (l-l_s)/N_\phi}
\Phi_{\{n_{\ga;l+l_s}\}}
\nonumber\\
&=\e^{\imth 2\pi \sum_{l=0}^{m-1} n_{\ga;l} (\frac{l}{m} +\frac{N_\phi-2l_s-m}{2m})}
\Phi_{\{n_{\ga;l+l_s}\}}
.
\end{align*}
If we choose
\begin{equation*}
 l_s=\frac{N_\phi-m+l_{max}}{2}
\end{equation*}
(which is always an integer since $N_\phi=mN_c$, $m=$ even, and
$l_{max}=S_n-S_{n-1}=$ even), we find that
\begin{align}
\label{T1T2Phi}
 T_1\Phi_\ga&=\e^{\imth 2\pi \sum_{l=0}^{m-1} (n_{\ga;l}-n_l) l/m} \Phi_\ga,
 =\e^{\imth 2\pi Q_\ga} \Phi_\ga,
\nonumber\\
 T_2\Phi_\ga&= \Phi_{\{n_{\ga;(l-1)\% m+l_s}\}} = \Phi_{\ga'}.
\end{align}
where we have used $\sum_{l=0}^{m-1} l n_l=nl_{max}/2$ (see \eq{nlsym}).  Here
$\ga'$ is the quasiparticle described by the canonical occupation distribution
$n_{\ga';l}=n_{\ga;(l-1)\% m}$.  We see that the eigenvalue of $T_1$ is related
to the charge of the corresponding quasiparticle $Q_\ga=\sum_{l=0}^{m-1}
(n_{\ga;l}-n_l) l/m$.  The action of $T_2$ just shifts the occupation
distribution by one step.  The new distribution describes a new quasiparticle
$\ga'$.

\subsubsection{Modular transformations}

The degenerate ground states on torus form a projective
representation of modular transformation\cite{Wrig} $M=\bpm a&b\\
c&d \epm \in SL(2,Z)$ , where $a,b,c,d \in Z$ and $ad-bc=1$ (see
appendix \ref{nabmodu}).  The modular representation may contain
information that completely characterize the topological order in
the corresponding FQH state.

The modular transformations are generated by
\begin{align}
 \label{MTMS}
M_T=\bpm 1&1\\ 0&1 \epm,\ \ \ \
M_S=\bpm 0&-1\\ 1&0 \epm.
\end{align}
For every $M\in SL(2,Z)$, we have an invertible transformation
$U(M)$ acting on the degenerate ground states on torus. $U(M)$'s
satisfy
\begin{equation*}
 U(M)U(M') \sim U(MM').
\end{equation*}
where $\sim$ mean equal up to a total phase factor. The two
generators $M_T$ and $M_S$ are represented as
\begin{equation}
\label{TSUM}
 T=U(M_T),\ \ \ \ \ \ S=U(M_S).
\end{equation}
$M=-1$ is represented as
\begin{equation*}
 C=U(-1)
\end{equation*}
which is the quasiparticle conjugation operator
(see \eq{CPhi}).

$T$, $S$, $T_1$ and $T_2$ have the following algebraic relation (see
appendix \ref{nabmodu})
\begin{align}
\label{TST1T2}
 TT_1&=T_1T, &  TT_2&=\e^{\imth \pi\frac{n}{m}} T_2T_1 T,
\nonumber\\
 ST_1&=T_2S, &  ST_2&= CT_1 C^{-1} S.
\end{align}
Since $M_S^2=-1$ and $(M_SM_T)^3=-1$, we also have
\begin{equation}
\label{ST23}
 S^2=C,\ \ \ \  C^2=1,\ \ \ \ (ST)^3 =\e^{\imth \th} C.
\end{equation}

\subsection{Quasiparticle tunneling around a torus}

\subsubsection{Quasiparticle tunneling operators}

To use the modular transformation to obtain the properties of
quasiparticles, it is useful to consider the following tunneling
process: (a) we first create a quasiparticle $\ga_1$ and its anti
quasiparticle $\ga_1^*$, then (b) move the quasiparticle around the
torus to wrap the torus $n_1$ times in the $X^1$ direction and $n_2$
times in the $X^2$ direction, and last (c) we annihilate $\ga_1$ and
$\ga_1^*$. The quasiparticle-pair creation process in the step (a)
is represented by an operator that map no-quasiparticle-particle
states to two-quasiparticle-particle states. The quasiparticle
transport process in the step (b) is represented by an operator that
map two-quasiparticle-particle states to two-quasiparticle-particle
states. The quasiparticle-pair annihilation process in the step (c)
is represented by an operator that map two-quasiparticle-particle
states to no-quasiparticle-particle states. The whole tunneling
process induces a transformation between the degenerate ground
states on the torus:
\begin{equation*}
 |\ga\>\to W^{(\ga_1)}_{(n_1,n_2)}|\ga\> = |\ga'\>
(W^{(\ga_1)}_{(n_1,n_2)})_{\ga'\ga}
\end{equation*}
For Abelian FQH states, $W^{(\ga_1)}_{(n_1,n_2)}$ is always an
invertible transformation.  But for non-Abelian FQH states,
$W^{(\ga_1)}_{(n_1,n_2)}$ may NOT be invertible. This is because
when we create the quasiparticle-anti-quasiparticle pair in the
step (a), the pair is in such a state that they fuse into the
identity channel.  But after wrapping the quasiparticle around the
torus, the fusion channel may change and hence the pair may not be
able to annihilate into ground in the step (c).  In other words,
the  annihilate process in the step (c) represents a projection
into the subspace spanned by the degenerate ground states.

\begin{figure}[t]
\centerline{
\includegraphics[scale=0.5]{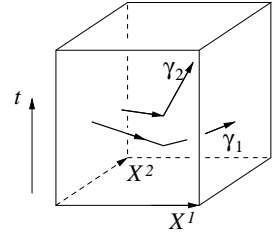}
} \caption{ The tunneling processes $A^{(\ga_1)}$ and $B^{(\ga_2)}$.
} \label{qptun}
\end{figure}

Let (see Fig. \ref{qptun})
\begin{equation*}
 A^{(\ga_1)}=W_{(1,0)}^{(\ga_1)},\ \ \ \ \
 B^{(\ga_1)}=W_{(0,1)}^{(\ga_1)}.
\end{equation*}
A combination of two tunneling processes in the $X^1$ direction:
$A^{(\ga_1)}$ and then $A^{(\ga_2)}$, induces a transformation
$A^{(\ga_2)}A^{(\ga_1)}$ on the degenerate ground states.  A
combination of the same two tunneling processes but with a different
time order: $A^{(\ga_2)}$ and then $A^{(\ga_1)}$, induces a
transformation $A^{(\ga_1)}A^{(\ga_2)}$ on the degenerate ground
states.  We note that the two tunneling paths with different time
orders can be deformed into each other smoothly. So they only differ
by local perturbations.  Due to the topological stability of the
degenerate ground states\cite{WNtop}, local perturbations cannot
change the degenerate ground states. Therefore $A^{(\ga_1)}$ and
$A^{(\ga_2)}$ commute, and similarly $B^{(\ga_1)}$ and $B^{(\ga_2)}$
commute too.
We see that $A^{(\ga)}$'s can be simultaneously diagonalized.
Similarly, $B^{(\ga)}$'s can also be simultaneously diagonalized.
Due to the $90^\circ$ rotation symmetry, $A^{(\ga)}$ and $B^{(\ga)}$
have the same set of eigenvalues.  But since $A^{(\ga)}$ and
$B^{(\ga)}$ in general do not commute, we in general cannot
simultaneously diagonalize $A^{(\ga)}$ and $B^{(\ga)}$.

The basis $\Phi_\ga=\Phi_{\{n_{\ga;l+l_s}\}}$ described by the
occupation distribution $n_{\ga;l+l_s}$ on the orbitals $\phi^{(l)}$
is a natural basis in which $A^{(\ga_1)}$ is diagonal.  This is
because the tunneling process $A^{(\ga_1)}$ does not move
quasiparticle in the $X^2$ direction, and hence does not modify the
occupation distribution $n_{\ga;l+l_s}$ on the orbitals
$\phi^{(l)}$.
On the other hand, $B^{(\ga_1)}$ does move quasiparticle in the
$X^2$ direction and hence shifts the occupation distribution
$n_{\ga;l+l_s}$ that characterizes the ground states.  Therefore
$B^{(\ga_1)}$ is not diagonal in the $\Phi_\ga$ basis. In
particular, when acted on the state $\Phi_0\equiv
\Phi_{\{n_{l+l_s}\}}$ that corresponds to the trivial
quasiparticle, $B^{(\ga_1)}$ produces the state $\Phi_{\ga_1}$
that corresponds the quasiparticle $\ga_1$:
\begin{equation}
\label{BPhiga}
 B^{(\ga_1)} \Phi_0=b_{\ga_1}\Phi_{\ga_1},
\end{equation}
where $b_{\ga_1}$ is a non-zero factor.  (Note that $\ga=0,1,\cdots,N_q-1$
where $\ga=0$ corresponds to the trivial quasiparticle and $N_q$ is the number
of quasiparticle types.)

\subsubsection{Tensor category structure in quasiparticle tunneling operators}

\begin{figure}[t]
\centerline{
\includegraphics[scale=0.5]{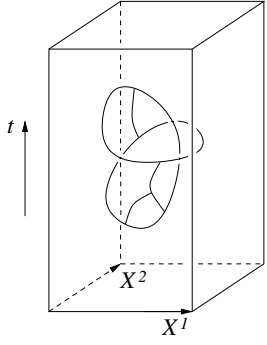}
}
\caption{
A general quasiparticle tunneling process.
}
\label{qptung}
\end{figure}

$W^{(\ga)}_{(n_1,n_2)}$ is just a special kind of quasiparticle
tunneling.  In general, we can create many pairs of quasiparticles,
move them around each other, combine and/or split quasiparticles,
and then annihilate all of them (see Fig. \ref{qptun}).  In addition
to the quasiparticle-pair creation process represented by a mapping
from no-quasiparticle-particle states to two-quasiparticle-particle
states and the quasiparticle-pair annihilation process represented
by a mapping from two-quasiparticle-particle states to
no-quasiparticle-particle states, the more general tunneling process
also contains the quasiparticle splitting process represented by a
mapping from one-quasiparticle-particle states to
two-quasiparticle-particle states and the quasiparticle fusion
process represented by a mapping from two-quasiparticle-particle
states to one-quasiparticle-particle states.  We will use $W(\cX)$
to represent the action of the whole tunneling process on the
degenerate ground state where $\cX$ represents the tunneling path. $
W^{(\ga_1)}_{(n_1,n_2)}$ discussed above is just a special case of
$W(\cX)$.

Due to the topological stability of the degenerate ground states
$W(\cX)$ can have some very nice algebraic properties.  However,
in order for $W(\cX)$ to have the nice algebraic structure we need
to choose the operators that represent the quasiparticle-pair
creation/annihilation and quasiparticle splitting/fusion processes
properly.  We conjecture that
after making those choices, $W(\cX)$ can satisfy the
following conditions:
\begin{align}
\label{Wrule}
W(\cX_\text{local}) &\propto 1 ,
\\
 W
\bpm \includegraphics[height=0.3in]{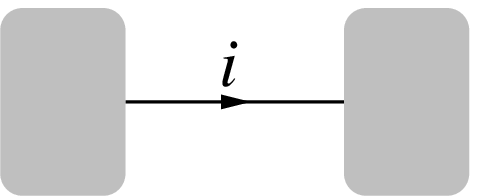} \epm  &=
W
\bpm \includegraphics[height=0.3in]{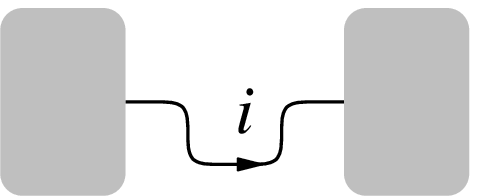} \epm ,
\nonumber\\
 W
\bpm \includegraphics[scale=0.40]{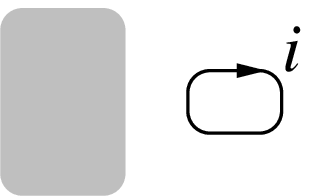} \epm  &=
d_iW
\bpm \includegraphics[scale=0.40]{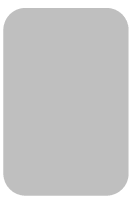} \epm ,
\nonumber\\
W
\bpm \includegraphics[scale=0.40]{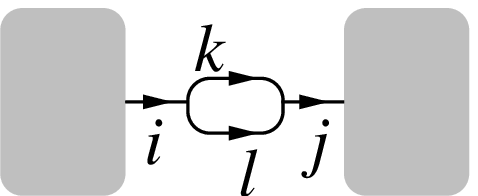} \epm  &=
\delta_{ij}
W
\bpm \includegraphics[scale=0.40]{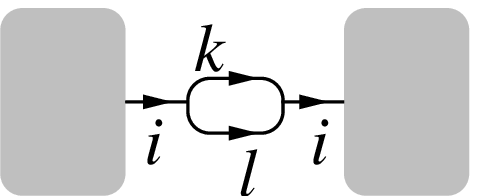} \epm ,
\nonumber\\
W
\bpm \includegraphics[scale=0.40]{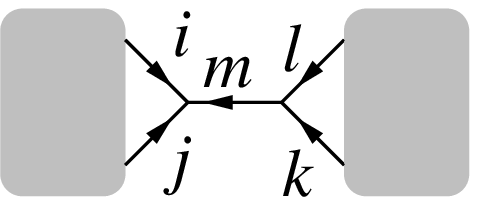} \epm  &=
\sum_{n=0}^{N_q-1}
F^{ijm}_{kln}
W
\bpm \includegraphics[scale=0.40]{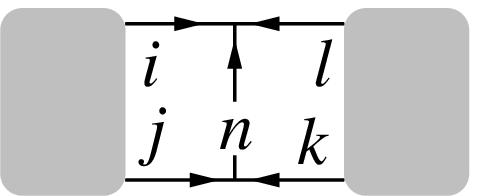} \epm  ,
\nonumber\\
W
\bpm \includegraphics[scale=0.35]{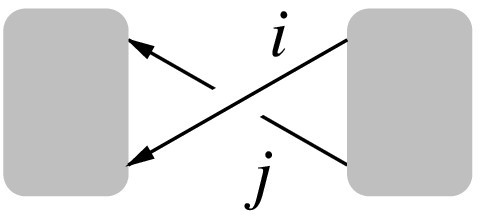}\epm &=
\sum_{k=0}^{N_q-1}
\om^{k}_{ij}
W
\bpm \includegraphics[scale=0.35]{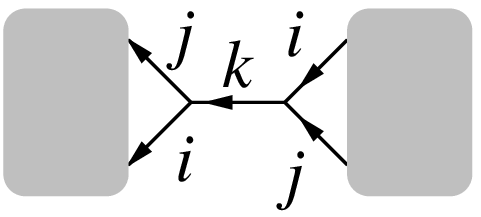}\epm
\nonumber\\
W
\bpm \includegraphics[scale=0.35]{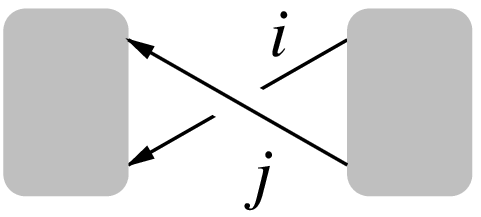}\epm &=
\sum_{k=0}^{N_q-1}
\om^{k}_{ij}
W\bpm\includegraphics[scale=0.35]{Brd2O.eps}\epm
\nonumber
\end{align}
where $i,j,\cdots=0,1,\cdots,N_q-1$ label the $N_q$ quasiparticle
types and $i=0$ corresponds to the trivial quasiparticle.  Note that
$N_q$ is also the ground state degeneracy on the torus.  The shaded
areas in \eqn{Wrule} represent other parts of tunneling path.  Note
that there may be tunneling paths that connect disconnected shaded
areas.

Here $\cX_\text{local}$ represents a tunneling path which has a
compact support (\ie there is no path in $\cX_\text{local}$ that
wraps around the torus). In this case $W(\cX_\text{local})$
represents local perturbations that cannot mix different degenerate
ground states on torus.\cite{WNtop} Thus $W(\cX_\text{local})$ must
be proportional to the identity.  The second relation in \eqn{Wrule}
implies that the tunneling amplitude $W(\cX)$ only depends on the
topology of the tunneling path. A smooth deformation of tunneling
path will not change $W(\cX)$.

Strictly speaking, due to the quasiparticle charge and the
external magnetic field, we only have $ W \bpm
\includegraphics[height=0.3in]{Xi0.eps} \epm  = \e^{\i \th} W \bpm
\includegraphics[height=0.3in]{Xis0.eps} \epm $ where $\th$ is a
path dependent phase.  However, we can restrict the tunneling
paths to be on a properly designed grid, such as a grid formed by
squares.  We choose the grid such that each square contains $k$
units of flux quanta where $k$ is an integer that satisfies
$kQ_\ga$= integer for any $\ga$.  In this case, $\e^{\i \th}=1$
for the tunneling paths on the grid.
Other relations in \eqn{Wrule} are motivated from tensor category
theory.\cite{BK,LWstrnet,K062}

One may notice that the rules \eq{Wrule} are about planar graphs
while the tunneling paths are three dimensional graphs. How can one
apply rules of planar graphs to three dimensional graphs? Here we
have picked a fixed direction of projection and projected the three
dimensional tunneling paths to a two dimensional plane.\cite{LWqed}
The rules \eq{Wrule} apply to such projected planar graphs.  Also,
the action of tunneling process can only be properly represented by
\emph{framed} graphs in three dimensions, to take into account the
phase factors generated by twisting the quasiparticles.  Here we
have assumed that there is a way to choose a canonical framing for
each tunneling process, such that their projections satisfy
\eqn{Wrule}.

The coefficients $(d_i,F^{ijm}_{kln},
\om^k_{ij})$ must satisfy the following self consistent
relations:\cite{LWstrnet,K062}
\begin{align}
\label{pentid}
F^{ijk}_{j^*i^*0} &= \frac{v_{k}}{v_{i}v_{j}} \del_{ijk} \nonumber \\
F^{ijm}_{kln} = F^{lkm^*}_{jin} &= F^{jim}_{lkn^*} = F^{imj}_{k^*nl}
\frac{v_{m}v_{n}}{v_{j}v_{l}} \nonumber \\
\sum_{n=0}^{N_q-1}
F^{mlq}_{kpn} F^{jip}_{mns} F^{jsn}_{lkr}
&= F^{jip}_{qkr} F^{riq}_{mls}
\nonumber\\
\omega^{m}_{js}F^{sl^*i}_{kjm^*}\omega^{l}_{si}
\frac{v_j v_s}{v_m} &=
\sum_{n=0}^{N}F^{ji^*k}_{s^*nl^*}\omega^{n}_{sk}F^{jl^*n}_{ksm^*}
\nonumber \\
\omega^{j}_{is} &= \sum_{k=0}^{N_q-1} \omega^{k}_{si^*}F^{i^*s^*k}_{isj*}
\end{align}
Here $i^*$ is the antiquasiparticle of $i$, $v_i$ is defined by
$v_i = v_{i^*} = \sqrt{d_i}$, and $\del_{ijk}$ is given by
\begin{equation*}
\del_{ijk}=
\begin{cases}
1, & \text{if $i,j,k$ can fuse into a trivial quasiparticle,}\\
0, & \text{otherwise}
\end{cases}
\end{equation*}
We would like to point out that here we only considered a quasiparticle fusion
algebra $\psi_i \psi_j=\sum_k N^k_{ij}\psi_k$ that has a special property
$N^k_{ij}=0,1$.

\subsubsection{Implications of tensor category structure}

\begin{figure}[t]
\centerline{
\includegraphics[scale=0.42]{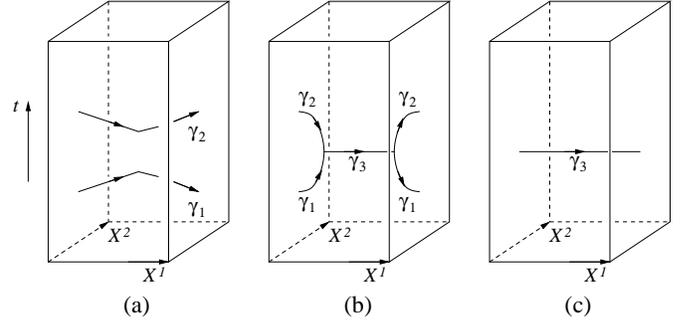}
} \caption{ (a): Two tunneling processes: $A^{(\ga_1)}$ and
$A^{(\ga_2)}$. (b): The tunneling path of the above two tunneling
processes can be deformed according to the fifth relation in
\eqn{Wrule}
with
$i=\ga_2^*$,
$j=\ga_2$,
$k=\ga_1$,
$k=\ga_1^*$,
$m=0$, and $n=\ga_3$.
(c): The fifth, the fourth, and the third relations in \eqn{Wrule} can
reduce
(b) to (c).
}
\label{falgT}
\end{figure}

As the first application of the above algebraic structure, we find that
(see Fig. \ref{falgT})
\begin{align}
\label{AAA}
 A^{(\ga_2)}
 A^{(\ga_1)}&=\sum_{\ga_3}
F^{\ga_2^*\ga_20}_{\ga_1\ga_1^*\ga_3}
d_{\ga_3} F^{\ga_3\ga_1^*\ga_2^*}_{\ga_2\ga_1^* 0}
 A^{(\ga_3)}
\nonumber\\
&=
\sum_{\ga_3} \del_{\ga_1\ga_2\ga_3^*} A^{(\ga_3)} .
\end{align}
We see that the algebra of $A^{(\ga)}$ forms a representation of fusion
algebra $\psi_{\ga_1}\psi_{\ga_2}=\sum_{\ga_3}
\del_{\ga_1\ga_2\ga_3^*}\psi_{\ga_3}$.  The operators
$B^{(\ga)}=SA^{(\ga)}S^{-1}$ (see \eqn{TSAB}) satisfy the same fusion algebra
\begin{align}
\label{BBB}
 B^{(\ga_2)} B^{(\ga_1)} &= \sum_{\ga_3} \del_{\ga_1\ga_2\ga_3^*} B^{(\ga_3)} .
\end{align}

\begin{figure}[t]
\centerline{
\includegraphics[scale=0.5]{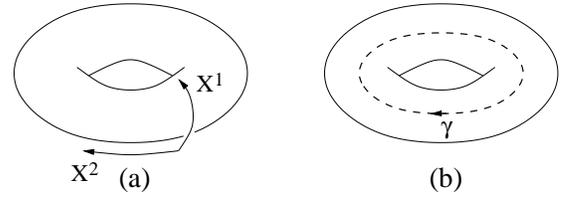}
}
\caption{
(a): The ground state $\Phi_{\ga=0}$ on a torus that corresponds to the trivial
quasiparticle can be represented by an empty solid torus.
(b): The other ground state $\Phi_{\ga}$ that corresponds to
a type $\ga$ quasiparticle can be represented by
an solid torus with a loop of type $\ga$ in the center.
}
\label{sttT}
\end{figure}

As the second application of the above algebraic structure, we can represent
the degenerate ground states on torus graphically.  One of the degenerate ground
state $\Phi_{0}$ that corresponds to the trivial quasiparticle $\ga=0$ can be
represented by an empty solid torus (see Fig. \ref{sttT}a).  We denote such a
state as $|0\>=\Phi_{0}$.  Other degenerated ground states can be obtained
by the action of the $B^{(\ga)}$ operators
\begin{equation}
\label{Bga0}
 |\ga\>\equiv B^{(\ga)} |0\> .
\end{equation}
From \eqn{BPhiga}, we see that $\Phi_\ga$ and $|\ga\>$ are related
\begin{equation*}
b_\ga \Phi_\ga=|\ga\> .
\end{equation*}
Since $|\ga\>$ is created by the tunneling operator $B^{(\ga)}$,
$|\ga\>$ can be represented by adding a loop that corresponds to the
$B^{(\ga)}$ operator to the center of the solid torus (see Fig.
\ref{sttT}b).

\begin{figure}[t]
\centerline{
\includegraphics[scale=0.5]{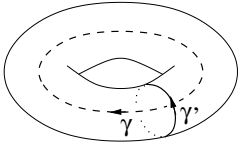}
}
\caption{
The graphic representation of $A^{(\ga')}|\ga\>$.
}
\label{sttTA}
\end{figure}

\begin{figure}[t]
\centerline{
\includegraphics[scale=0.35]{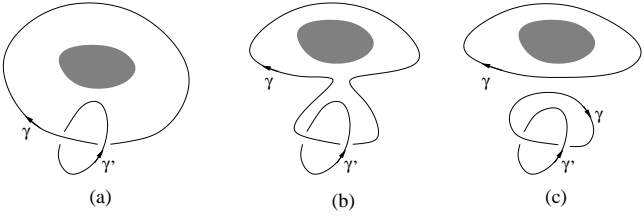}
}
\caption{
(a): The graphic representation of $A^{(\ga')}|\ga\>$.  (b):  The graphic
representation of $A^{(\ga')}|\ga\>$.  (c): The fifth and the fourth relations
in \eqn{Wrule} can deformed the graph in (b) to the graph in (c).
The shaded area represents the hole of the torus.
}
\label{sttTAlnk}
\end{figure}

$|\ga\>$ is a natural basis for tensor category theory.  The matrix elements
of $A^{(\ga)}$ and $B^{(\ga)}$ have simple forms in such a basis.  From
\eqn{BBB}, we see that
\begin{align*}
B^{(\ga_2)} B^{(\ga_1)}|0\> &=
B^{(\ga_2)} |\ga_1\>  = \sum_{\ga_3} \del_{\ga_1\ga_2\ga_3^*} |\ga_3\> .
\end{align*}
Therefore, in the basis $|\ga\>$, the matrix elements of $B^{(\ga_2)}$ are
given by the coefficients of fusion algebra
\begin{equation}
\label{Bdel}
 B^{(\ga_2)}_{\ga_3\ga_1}=\del_{\ga_1\ga_2\ga_3^*} .
\end{equation}
The action of $A^{(\ga')}$ on $|\ga\>$ is represented by Fig. \ref{sttTA}.
From Fig. \ref{sttTAlnk}, we find that
\begin{equation}
\label{AgaS}
 A^{(\ga')}|\ga\>=\frac{S^\text{TC}_{\ga^*\ga'}}{d_\ga} |\ga\> ,
\end{equation}

We see that $A^{(\ga')}$ is diagonal in the $|\ga\>$ basis.  Let
$a^{(\ga')}_\ga$ be the eigenvalues of $A^{(\ga')}$ we see that
\begin{equation}
\label{aSTC}
 a^{(\ga')}_\ga=\frac{S^\text{TC}_{\ga^*\ga'}}{d_\ga}
=\frac{S^\text{TC}_{\ga'\ga^*}}{S^\text{TC}_{0\ga^*}} .
\end{equation}

\begin{figure}[t]
\centerline{
\includegraphics[scale=0.5]{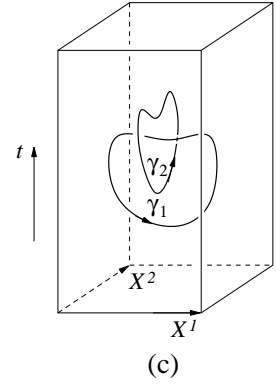}
}
\caption{
The amplitude of two linked local loops
is a complex number $S^\text{TC}_{\ga_1\ga_2}$.
}
\label{lnklp}
\end{figure}

$S^\text{TC}_{\ga_1\ga_2}$ in \eqn{AgaS} is the amplitude of two linked local loops (see
Fig. \ref{lnklp}).
$S^\text{TC}_{\ga_1\ga_2}$ satisfies
\begin{equation}
\label{STCsymm}
 S^\text{TC}_{\ga_1\ga_2} =S^\text{TC}_{\ga_2\ga_1}
  .
\end{equation}
Using the tensor category theory \eq{Wrule}, one can also show that\cite{K062}
\begin{align}
\label{SSSdel}
 \frac{S^\text{TC}_{\ga_1\ga} S^\text{TC}_{\ga_2\ga} }{ S^\text{TC}_{0\ga} }
=\sum_{\ga_3} \del_{\ga_1\ga_2\ga^*_3} S^\text{TC}_{\ga_3\ga}
=\sum_{\ga_3} S^\text{TC}_{\ga\ga_3} B^{(\ga_1)}_{\ga_3\ga_2}
\end{align}
which can be rewritten as
\begin{equation}
\label{SSA}
 \frac{S^\text{TC}_{\ga_1\ga}}{ S^\text{TC}_{0\ga} }\del_{\ga\ga'}
 =\sum_{\ga_3\ga_2} S^\text{TC}_{\ga\ga_3} B^{(\ga_1)}_{\ga_3\ga_2} ((S^\text{TC})^{-1})_{\ga_2\ga'}
=A^{(\ga_1)}_{\ga^*\ga^{\prime *}} .
\end{equation}
In the operator form, the above becomes
\begin{equation}
\label{ACsB}
 A^{(\ga_1)} = C S^\text{TC}B^{(\ga_1)}(S^\text{TC})^{-1} C,
\end{equation}
where $C$ is the charge conjugation operator $ C|\ga\>=|\ga^*\>$. We
see that $S^\text{TC}$ can change the tunneling operator
$B^{(\ga_1)}$ to $A^{(\ga_1)}$. \Eqn{SSSdel} can also be rewritten
as (assume $S^\text{TC}$ is invertible)
\begin{align}
\label{SSSdel1}
\sum_\ga
 \frac{
S^\text{TC}_{\ga_1\ga} S^\text{TC}_{\ga_2\ga}
((S^\text{TC})^{-1})_{\ga \ga_3}
}{ S^\text{TC}_{0\ga} }
= \del_{\ga_1\ga_2\ga^*_3}
\end{align}
We note that the above expression is invariant under
$S^\text{TC}_{\ga'\ga} \to S^\text{TC}_{\ga'\ga} f_\ga $.

$T_1$ and $T_2$ generate the translation symmetry of the torus.  We expect
that $T_1$ and $T_2$ commute with the algebraic structure of the tensor
category. Thus we expect that \eqn{T1T2Phi} keeps the same form in the new
basis $|\ga\>$:
\begin{equation}
\label{T1T2ga}
T_1|\ga\>=\e^{\i 2\pi Q_\ga} |\ga\>,\ \ \ \ \ \ \ \
T_2|\ga\>=|\ga'\>,
\end{equation}
where $\ga'$ is the quasiparticle described by the canonical occupation
distribution $n_{\ga';l}=n_{\ga;(l-1)\% m}$.

\subsubsection{Quasiparticle tunneling operators under modular transformation}

The transformations $W_{(n_1,n_2)}^{(\ga)}$ induced by quasiparticle
tunneling processes have certain algebraic relation with the modular
transformations $U(M)$.  From Fig. \ref{modutran}, we see that the
modular transformation $M_T$ changes $W_{(n_1,n_2)}^{(\ga)}$ to
$W_{(n_1+n_2,n_2)}^{(\ga)}$:
\begin{align}
\label{TWn1n2}
 T W_{(n_1,n_2)}^{(\ga_1)} &= W_{(n_1+n_2,n_2)}^{(\ga_1)} T   .
\end{align}
Since the modular transformation $M_S$ generates a $90^\circ$
rotation, we find
\begin{align}
\label{SWn1n2}
 S W_{(n_1,n_2)}^{(\ga_1)} &= W_{(-n_2,n_1)}^{(\ga_1)} S .
\end{align}
Here $M_S$ and $M_S$ are given by \eq{MTMS} and $T$ and $S$ are
given by \eq{TSUM}.  Also since the modular transformation $M=-1$
generates a $180^\circ$ rotation, we find
\begin{align}
\label{CWn1n2}
 C W_{(n_1,n_2)}^{(\ga_1)} &= W_{(-n_2,-n_1)}^{(\ga_1)} C .
\end{align}
In terms of $A^{(\ga_1)}$ and $B^{(\ga_1)}$
we can rewrite \eqn{TWn1n2}, \eqn{SWn1n2}, and \eqn{CWn1n2} as
\begin{align}
\label{TSAB}
 T A^{(\ga_1)} &= A^{(\ga_1)} T   ,
\nonumber\\
 T B^{(\ga_1)} &= W_{(1,1)}^{(\ga_1)}T ,
\nonumber\\
 S A^{(\ga_1)} &= B^{(\ga_1)} S ,
\nonumber\\
S B^{(\ga_1)} C &= C A^{(\ga_1)} S ,
.
\end{align}

Let us assume that the set of the quasiparticle operators $A^{(\ga)}$ can
resolve all the degenerate ground states $|\ga\>$, \ie no two degenerate ground
states share the common set of eigenvalues for  the operators $A^{(\ga)}$. In
this case, the commutation relation $T A^{(\ga_1)} = A^{(\ga_1)} T$ implies
that $T$ is diagonal in the $|\ga\>$ basis.  We will fix the over all phase
factor of $T$ by choosing $T_{00}=1$.

The operator $C=S^2$ is a charge conjugation operator. Its action
on $|\ga\>$ is given by (see \eqn{nbarganga})
\begin{equation*}
 C|\ga\>=|\ga^*\> .
\end{equation*}

Compare $A^{(\ga_1)}  = S^{-1}B^{(\ga_1)} S=CSB^{(\ga_1)} S^{-1}C$ with
\eq{ACsB}, we find that
$S^\text{TC}= S F$ where $F$ is a diagonal matrix
$ F_{\ga\ga'}=f_\ga\del_{\ga\ga'} $
in the $|\ga\>$ basis.
Using $S^\text{TC}= S F$, we can rewrite
\eqn{aSTC} and \eqn{SSSdel1} as
\begin{equation}
\label{Sba}
\frac{
S_{\ga_1\ga}
f_\ga
}{
S_{0\ga}
f_\ga
}
=
\frac{
S_{\ga_1\ga} }{
S_{0\ga}
}
= a^{(\ga_1)}_{\ga^*}  ,
\end{equation}
\begin{align}
\label{SSSdel2}
\sum_\ga
 \frac{
S_{\ga_1\ga} S_{\ga_2\ga}
(S^{-1})_{\ga \ga_3}
}{ S_{0\ga} }
= \del_{\ga_1\ga_2\ga^*_3}
\end{align}
We see that $A^{(\ga)}$ and $B^{(\ga)}$ can be determined from $S$.


Since $S^\text{TC}$ is symmetric, $S^\text{TC}_{0\ga}=d_\ga>0$
and $S^\text{TC}_{00}=1$,
once we know $S$, we can use those
conditions to fix $F$. Thus we can determine $S^\text{TC}$ from $S$.
Once we know $S^\text{TC}$, we can also calculate
the CFT scaling dimension $h_\ga$ for the quasiparticle $\ga$
(see appendix \ref{cftsec})
up to an integer:\cite{K062}
\begin{equation}
\label{Shhh}
S^\text{TC}_{\ga_1\ga_2}=\sum_{\ga_3}
\del_{\ga_1\ga_2\ga_3^*}\e^{\i 2\pi(h_{\ga_3}-h_{\ga_1}-h_{\ga_2} )} d_{\ga_3}
.
\end{equation}
The CFT scaling dimensions for the quasiparticles
determine the $T$ matrix:
\begin{equation}
\label{Thga}
 T_{\ga\ga'}=\e^{\i 2\pi h_{\ga} } \del_{\ga\ga'} .
\end{equation}

\subsection{Summary}

In this section, we calculated the charge and the orbital spin of
quasiparticle, as well as and the ground state degeneracy from the
pattern of zeros $\{S_a\}$ of a FQH states.  We also discussed the
translation transformations, the modular transformations, and the
transformations induced by the quasiparticle tunneling on the
degenerate ground states.  The algebra of those transformation can
help us to determine the quasiparticle statistics, quasiparticle
quantum dimensions, and fusion algebra of the quasiparticles. In
particular, we can use the algebra \eq{TST1T2} and \eq{ST23} to
determine $S$, and then use \eqn{Sba}, \eqn{SSSdel2}, and \eqn{Shhh} to
determine the quasiparticle tunneling operators, $A^{(\ga)}$ and
$B^{(\ga)}$, and quasiparticle scaling dimensions $h_\ga$.  The
condition that the matrix elements of $B^{(\ga)}$ must be
non-negative integers put further constraint on $S$.

\section{Examples}
\label{exmpl}

\subsection{Quasiparticles in FQH states}

In \Ref{WWsymm}, many FQH states are characterized and constructed through
patterns of zeros.  The pattern of zeros in FQH states can be characterized by
a $\v S$-vector $\v S= (m;S_2,\cdots,S_n)$, or a $\v h$-vector $\v h=
(\frac{m}{n}; h^\text{sc}_1,\cdots,h^\text{sc}_n)$, or an occupation
distribution $n_0\cdots n_{m-1})$.  All those data contain information on two
important integers $n$ and $m$.  $n$ is the number of electrons in one cluster
and $m$ determines the filling fraction $\nu=n/m$.

The FQH states constructed in \Ref{WWsymm} include the $Z_n$ parafermion
states $\Phi_{\frac{n}{m};Z_n}$ introduced in \Ref{MR9162,RR9984}.  The
patterns of zeros $\{S_a\}$ for those $Z_n$ parafermion states are obtained.
The occupation distributions $\{n_l\}$  of those states agree with those
obtained in \Ref{SL0604,BKW0608,ABK0875}.  \Ref{WWsymm} also obtained
generalized $Z_n$ parafermion states $\Phi_{\frac{n}{m};Z_n^{(k)}}$ and their
patterns of zeros.  $\Phi_{\frac{n}{m};Z_n^{(k)}}$ has a filling fraction
$\nu=n/m$.  Many other new FQH states and their patterns of zeros are also
obtained in \Ref{WWsymm}, such as the $\Phi_{\frac{n}{2n};C_n}$ and the
$\Phi_{\frac{n}{n};C_n/Z_2}$ states.

Once we know the pattern of zeros of a FQH state, we can find all
the quasiparticle excitations in such a state, by simply finding
all $S_{\ga,a}$ that satisfy \eq{SqCond} and \eq{ngalcan} (note
that $n_{\ga;l}$ in \eq{ngalcan} are determined from $S_{\ga;a}$).
From the pattern of zeros that characterizes a quasiparticles, we
can find many quantum numbers of that quasiparticle.  Here we will
summarize those results by just listing the number of
quasiparticle types in some FQH states. Then, we will give a more
detailed discussion for few simple examples.

\begin{widetext}

For the parafermion states $\Phi_{\frac{n}{2};Z_n}$ ($m=2$), we find
the numbers of quasiparticle types (NOQT) to be
\begin{align*}
 \begin{tabular}{|r|c|c|c|c|c|c|c|c|c|}
\hline
FQH state: &
$\Phi_{\frac{2}{2};Z_2}$ &
$\Phi_{\frac{3}{2};Z_3}$ &
$\Phi_{\frac{4}{2};Z_4}$ &
$\Phi_{\frac{5}{2};Z_5}$ &
$\Phi_{\frac{6}{2};Z_6}$ &
$\Phi_{\frac{7}{2};Z_7}$ &
$\Phi_{\frac{8}{2};Z_8}$ &
$\Phi_{\frac{9}{2};Z_9}$ &
$\Phi_{\frac{10}{2};Z_{10}}$
\\
\hline
NOQT: &
3  &
4  &
5  &
6  &
7  &
8  &
9  &
10  &
11
\\
\hline
 \end{tabular}
\end{align*}
For the parafermion states $\Phi_{\frac{n}{2+2n};Z_n}
=\Phi_{\frac{n}{2};Z_n}\prod_{i<j}(z_i-z_j)^2 $ ($m=2+2n$), we find
\begin{align*}
 \begin{tabular}{|r|c|c|c|c|c|c|c|c|c|}
\hline
FQH state: &
$\Phi_{\frac{2}{6};Z_2}$ &
$\Phi_{\frac{3}{8};Z_3}$ &
$\Phi_{\frac{4}{10};Z_4}$ &
$\Phi_{\frac{5}{12};Z_5}$ &
$\Phi_{\frac{6}{14};Z_6}$ &
$\Phi_{\frac{7}{16};Z_7}$ &
$\Phi_{\frac{8}{18};Z_8}$ &
$\Phi_{\frac{9}{20};Z_9}$ &
$\Phi_{\frac{10}{22};Z_{10}}$
\\
\hline
NOQT: &
9  &
16  &
25  &
36  &
49  &
64  &
81  &
100  &
121
\\
\hline
 \end{tabular}
\end{align*}
For the generalized parafermion states $\Phi_{\frac{n}{m};Z_n^{(k)}}$, we find
\begin{align*}
 \begin{tabular}{|r|c|c|c|c|c|c|c|c|c|}
\hline
FQH state: &
$\Phi_{\frac{5}{8};Z_5^{(2)}}$ &
$\Phi_{\frac{5}{18};Z_5^{(2)}}$ &
$\Phi_{\frac{7}{8};Z_7^{(2)}}$ &
$\Phi_{\frac{7}{22};Z_7^{(2)}}$ &
$\Phi_{\frac{7}{18};Z_7^{(3)}}$ &
$\Phi_{\frac{7}{32};Z_7^{(3)}}$ &
$\Phi_{\frac{8}{18};Z_8^{(3)}}$ &
$\Phi_{\frac{9}{8};Z_9^{(2)}}$ &
$\Phi_{\frac{10}{18};Z_{10}^{(3)}}$
\\
\hline
NOQT: &
24  &
54  &
32  &
88  &
72  &
128  &
81  &
40  &
99
\\
\hline
 \end{tabular}
\end{align*}
where $k$ and $n$ are coprime.
For the composite parafermion states
$
\Phi_{\frac{n_1}{m_1};Z_{n_1}^{(k_2)}}
\Phi_{\frac{n_2}{m_2};Z_{n_2}^{(k_2)}}
$ obtained as products of two parafermion wave functions , we find
\begin{align*}
 \begin{tabular}{|r|c|c|c|c|}
\hline
FQH state: &
$ \Phi_{\frac{2}{2};Z_2} \Phi_{\frac{3}{2};Z_3} $ &
$ \Phi_{\frac{3}{2};Z_3} \Phi_{\frac{4}{2};Z_4} $ &
$ \Phi_{\frac{2}{2};Z_2} \Phi_{\frac{5}{2};Z_5} $ &
$ \Phi_{\frac{2}{2};Z_2} \Phi_{\frac{5}{8};Z_5^{(2)}} $
\\
\hline
NOQT: &
30  &
70  &
63  &
117
\\
\hline
 \end{tabular}
\end{align*}
where $n_1$ and $n_2$ are coprime.  The filling fractions of the above
composite states are $\nu=\frac{n_1 n_2}{m_1 n_2+m_2 n_1}$.
\end{widetext}

The above results suggest a pattern.
For a (generalized) parafermion state $\Phi_{\frac{n}{m};Z_n^{(k)}}$,
we can express its filling fraction as $\nu=n/m=p/q$ where $p$ and $q$
are coprime. Then the number of quasiparticle types is given by
NOQT$=qD(n)$ where
$D(2)=3$,
$D(3)=2$,
$D(4)=5$,
$D(5)=3$,
$D(6)=7$,
$D(7)=4$,
$D(8)=9$,
$D(9)=5$, and
$D(10)=11$; or $D(n)=n+1$ for $n=$ even and $D(n)=\frac{n+1}{2}$
for $n=$ odd.
Similarly, For a composite parafermion state
$
\Phi_{\frac{n_1}{m_1};Z_{n_1}^{(k_1)}}
\Phi_{\frac{n_2}{m_2};Z_{n_2}^{(k_2)}}
$, we can express its
filling fraction as $\nu=\frac{n_1 n_2}{m_1 n_2+m_2 n_1}=p/q$ where $p$ and $q$
are coprime. Then the number of quasiparticle types is given by
NOQT$=qD(n_1)D(n_2)$.

The corresponding CFT of the above (generalized and composite) parafermion
states are known.  The numbers of the quasiparticle types can also be
calculated from the CFT.\cite{BW}
For the generalized parafermion state
$\Phi_{\frac{n}{m};Z_n^{(k)}}$ the  numbers for the quasiparticle types
is given by\cite{BW}
\begin{equation*}
 NOQT=\frac{1}{\nu}\frac{n(n+1)}{2} = \frac{m}{n}\frac{n(n+1)}{2}=
\frac{m(n+1)}{2} .
\end{equation*}
For the composite parafermion state
$\prod_i \Phi_{\frac{n_i}{m_i};Z_{n_i}^{(k_i)}}$ the
numbers for the quasiparticle types
is given by\cite{BW}
\begin{equation*}
 NOQT=\frac{1}{\nu}\prod_i \frac{n_i(n_i+1)}{2} .
\end{equation*}
Here we require that $k_i$ is not a factor of $n_i$
and $n_1$, $n_2$, $n_3$, $\cdots$ have no common factor.
The CFT approach gives rise to exactly the same
numbers for the quasiparticle types.

\begin{widetext}

For generalized parafermion states
$
\Phi_{\frac{n}{m};Z_{n}^{(k)}}
$ where $n$ and $k$ have a common factor, we have
\begin{align*}
 \begin{tabular}{|r|c|c|c|c|c|c|}
\hline
FQH state: &
$ \Phi_{\frac{4}{8};Z_4^{(2)}}$ &
$ \Phi_{\frac{6}{8};Z_6^{(2)}}$ &
$ \Phi_{\frac{6}{8};Z_6^{(3)}}$ &
$ \Phi_{\frac{8}{8};Z_8^{(2)}}$ &
$ \Phi_{\frac{8}{8};Z_8^{(4)}}$ &
$ \Phi_{\frac{9}{18};Z_9^{(3)}}$
\\
\hline
NOQT: &
10  &
20  &
21  &
35 &
36 &
56
\\
\hline
 \end{tabular}
\end{align*}
For more general composite parafermion states
$
\Phi_{\frac{n_1}{m_1};Z_{n_1}^{(k_2)}}
\Phi_{\frac{n_2}{m_2};Z_{n_2}^{(k_2)}}
$ where $n_1$ and $n_2$ have a common factor, we have
\begin{align*}
 \begin{tabular}{|r|c|c|c|c|c|c|c|}
\hline
FQH state: &
$ \Phi_{\frac{2}{2};Z_2} \Phi_{\frac{2}{2};Z_2} $ &
$ \Phi_{\frac{2}{2};Z_2} \Phi_{\frac{4}{2};Z_4} $ &
$ \Phi_{\frac{3}{2};Z_3} \Phi_{\frac{3}{2};Z_3} $ &
$ \Phi_{\frac{4}{2};Z_4} \Phi_{\frac{4}{2};Z_4} $  &
$ \Phi_{\frac{5}{2};Z_5} \Phi_{\frac{5}{2};Z_5} $   &
$ \Phi_{\frac{5}{2};Z_5} \Phi_{\frac{5}{8};Z_5^{(2)}} $   &
$ \Phi_{\frac{5}{8};Z_5^{(2)}} \Phi_{\frac{5}{8};Z_5^{(2)}} $
\\
\hline
NOQT: &
10  &
42  &
20  &
35  &
56&
352&
224
\\
\hline
 \end{tabular}
\end{align*}
For the $\nu=1/2$ states $\Phi_{\frac{n}{2n};C_n}$, we have
\begin{align*}
 \begin{tabular}{|r|c|c|c|c|c|c|c|c|}
\hline
FQH state: &
$\Phi_{\frac{3}{6};C_3}$ &
$\Phi_{\frac{4}{8};C_4}$ &
$\Phi_{\frac{5}{10};C_5}$ &
$\Phi_{\frac{6}{12};C_6}$ &
$\Phi_{\frac{7}{14};C_7}$ &
$\Phi_{\frac{8}{18};C_8}$ &
$\Phi_{\frac{9}{18};C_9}$
\\
\hline
NOQT: &
56  &
170  &
352  &
910  &
1612  &
3546  &
6266
\\
\hline
 \end{tabular}
\end{align*}
Note that $\Phi_{\frac{5}{10};C_5}$ and
$ \Phi_{\frac{5}{2};Z_5} \Phi_{\frac{5}{8};Z_5^{(2)}} $
have the same pattern of zeros and may be the same state.
For the $\nu=1$ states $\Phi_{\frac{n}{n};C_n/Z_2}$, we have
\begin{align*}
 \begin{tabular}{|r|c|c|c|c|c|c|c|c|}
\hline
FQH state: &
$\Phi_{\frac{4}{4};C_4/Z_2}$ &
$\Phi_{\frac{6}{6};C_6/Z_2}$ &
$\Phi_{\frac{8}{8};C_8/Z_2}$ &
$\Phi_{\frac{10}{10};C_{10}/Z_2}$
\\
\hline
NOQT: &
35  &
138  &
171  &
338
\\
\hline
 \end{tabular}
\end{align*}
Note that
$\Phi_{\frac{8}{8};Z_8^{(2)}}$,
$\Phi_{\frac{4}{4};C_4/Z_2}$ and
$ \Phi_{\frac{4}{2};Z_4} \Phi_{\frac{4}{2};Z_4} $
have the same pattern of zeros and may be the same state.

\end{widetext}

For those more general and new FQH states, the corresponding CFT are not
identified. Even the stability of those FQH states is unclear. If some of
those states contain gapless excitations, then the number of quasiparticle
types will make no sense for those gapless states.  In the following, we will
study a few simple examples in more details.

\subsection{$\nu=1/2$ Laughlin state}

For the $\nu=1/2$ Laughlin state, $n=1$ and its
pattern of zeros is characterized by
\begin{align*}
\Phi_{1/2}:\  (m;S_2,\cdots,S_n) &=(2;),
\nonumber\\
(\frac{m}{n}; h^\text{sc}_1,\cdots,h^\text{sc}_n) &=(\frac{2}{1}; 0),
\nonumber\\
n_0\cdots n_{m-1}&=10
\end{align*}

Solving \eq{SqCond}, we find that
there are two types of quasiparticles. Their canonical
occupation distributions and other quantum numbers are given by
\begin{align*}
 \begin{tabular}{|c|c|c|}
\hline
$n_{\ga;l}$ & $Q_\ga$ &  $S^\text{osp}_\ga$ \\
\hline
10 & 0        & 0       \\
01 & 1/2      & 1/4     \\
\hline
 \end{tabular}
\end{align*}
Thus $T_1$, $T_2$, and $T$ are given by
\begin{equation*}
 T_1=
\bpm
1 & 0\\
0& -1
\epm,\ \ \ \
 T_2=
\bpm
0 & 1\\
1 & 0
\epm,\ \ \ \
 T=
\bpm
1 & 0\\
0 & t_1
\epm  .
\end{equation*}
Since $l_{max}=0$, from \eq{nbarganga}, we find that
the conjugate of $10$ is $\overline{10}=10$ and
the conjugate of $01$ is $\overline{01}=01$.
Thus $ C=1$.

From $TT_2=\e^{\i \pi n/m} T_2T_1 T$ (see \eqn{TST1T2})
we find that $t_1=\i$.
From $ST_1=T_2S$, we find that
$S_{10}=S_{00}$ and $S_{01}=-S_{11}$.
From $S^2=C=1$, we find that $S_{11}=-S_{00}$,
and $S_{00}=\pm \frac{1}{\sqrt{2}}$.
Thus
\begin{equation*}
 S=\frac{1}{\sqrt{2}}
\bpm
1 & 1 \\
1 & -1
\epm ,\ \ \ \ \
 T=
\bpm
1 & 0\\
0 & \i
\epm   .
\end{equation*}
The above implies $(ST)^3=\e^{\imth \pi/4}$.  Those modular
transformations, $S$ and $T$, agree with those calculated using the
Chern-Simons effective theory\cite{Wrig} and the explicit FQH wave
functions.\cite{KW9327} From $T_{11}=\i$, we find that the
quasiparticle $(n_{\ga;l})=(01)$ has a scaling dimension $h=1/4$ and
a semion statistics.

Let us introduce
\begin{equation*}
 T_m=\e^{\imth 2\pi/24} T,\ \ \ \ \ S_m=S,
\end{equation*}
We find that
$T_m$ and $S_m$ satisfy
\begin{equation*}
 S_m=1,\ \ \ \ (S_mT_m)^3=1.
\end{equation*}
Thus $S_m$ and $T_m$ generate a linear representation of the modular group.

\subsection{$Z_2$ parafermion state}

The $\nu=1$ bosonic Pfaffian state\cite{MR9162} is
a $Z_2$ parafermion state with $n=2$.
Its pattern of zeros is described by
\begin{align*}
\Phi_{\frac{2}{2};Z_2}:\  (m;S_2,\cdots,S_n) &=(2;0),
\nonumber\\
(\frac{m}{n}; h^\text{sc}_1,\cdots,h^\text{sc}_n) &=(\frac{2}{2}; \frac12,0),
\nonumber\\
(n_0,\cdots,n_{m-1})&=(2,0)
\end{align*}
Solving \eq{SqCond}, we find that
there are three types of quasiparticles:
\begin{align*}
 \begin{tabular}{|c|c|c|}
\hline
$n_{\ga;l}$ & $Q_\ga$ & $S^\text{osp}_\ga$  \\
\hline
20 & 0      & 0       \\
02 & 1      & 1/2     \\
11 & 1/2    & 1/4     \\
\hline
 \end{tabular}
\end{align*}
Thus in the $|\ga\>$ basis, $T_1$, $T_2$ and $T$ are given by
\begin{equation*}
 T_1=
\bpm
1 & 0 & 0\\
0& 1 & 0\\
0&  0 & -1\\
\epm,\ \ \ \
 T_2=
\bpm
0 & 1 & 0\\
1 & 0 & 0\\
0 & 0 & 1\\
\epm,\ \ \ \
 T=
\bpm
1 & 0 & 0\\
0& t_1 & 0\\
0&  0 & t_2\\
\epm  .
\end{equation*}
Since $l_{max}=0$, from \eq{nbarganga}, we find that $\overline{20}=20$,
$\overline{02}=02$, and $\overline{11}=11$.  Thus $ C=1 $.

From $TT_2=\e^{\i \pi n/m} T_2T_1 T$ (see \eqn{TST1T2})
we find that $t_1=-1$, but $t_2$ is undetermined.
Using $ST_1=T_2S$, $ST_2=T_1S$, $S^T=S$, and $S^2=1$,
we  obtain the following possible solutions:
\begin{equation*}
  S=
\frac12 \bpm
1 & 1 & \sqrt 2 x\\
1&  1 & -\sqrt 2 x\\
\sqrt 2/x&  -\sqrt 2/x & 0\\
\epm  .
\end{equation*}
Those $S$'s satisfy $(ST)^3=  t_2 $.

Using the above $S$, we can calculate the fusion coefficients
$\del_{\ga_1\ga_2\ga_3}$ and $B^{(\ga)}_{\ga_1\ga_2}$ from \eqn{SSSdel2}.  We
find that $\del_{220}=\del_{221}=1/x^2$.  The condition $\del_{\ga\ga^*0}=1$
fixes $x=\pm 1$.  Thus we  have the following two possible solutions:
\begin{equation*}
  S_1=
\frac12 \bpm
1 & 1 & \sqrt 2 \\
1&  1 & -\sqrt 2 \\
\sqrt 2&  -\sqrt 2 & 0\\
\epm  ,
\end{equation*}
and
\begin{equation*}
  S_2=
\frac12 \bpm
1 & 1 & -\sqrt 2 \\
1&  1 & \sqrt 2 \\
-\sqrt 2&  \sqrt 2 & 0\\
\epm  .
\end{equation*}
We note that the above $S$'s are already symmetric.
Thus those $S$'s can regarded as $S^\text{TC}$.

Now let us calculate $A^{(\ga)}$ and $B^{(\ga)}$, $\ga=0,1,2$
using \eqn{Sba}. For the first solution $S_1$ we find that
\begin{align*}
 (a^{(\ga')}_\ga)=
\bpm
1&1&\sqrt 2 \\
1&1&-\sqrt 2 \\
1&-1&0\\
\epm
\end{align*}
where $\ga$ labels rows and $\ga'$ labels columns.
We have
\begin{align*}
 A^{(0)}&=
\bpm
1&0&0\\
0&1&0\\
0&0&1\\
\epm,\ \ \ \ \
 A^{(1)}=
\bpm
1&0&0\\
0&1&0\\
0&0&-1\\
\epm,
\nonumber\\
 A^{(2)} &=
\bpm
\sqrt 2&0&0\\
0&-\sqrt 2&0\\
0&0&0\\
\epm  .
\end{align*}
We see that
\begin{align*}
\bpm
A^{(0)} A^{(0)} & A^{(0)} A^{(1)} & A^{(0)} A^{(2)} \\
A^{(1)} A^{(0)} & A^{(1)} A^{(1)} & A^{(1)} A^{(2)} \\
A^{(2)} A^{(0)} & A^{(2)} A^{(1)} & A^{(2)} A^{(2)} \\
\epm
=
\bpm
A^{(0)} &  A^{(1)} & A^{(2)} \\
A^{(1)} &  A^{(0)} & A^{(2)} \\
A^{(2)} &  A^{(2)} & A^{(0)}+A^{(1)} \\
\epm .
\end{align*}
We recover the fusion algebra of the
$Z_2$ parafermion theory.
We also find that
\begin{align*}
 B^{(0)}&=
\bpm
1&0&0\\
0&1&0\\
0&0&1\\
\epm,\ \ \ \ \
 B^{(1)}=
\bpm
0&1&0\\
1&0&0\\
0&0&1\\
\epm,
\nonumber\\
 B^{(2)} &=
\bpm
0&0&1\\
0&0&1\\
1&1&0\\
\epm  .
\end{align*}
We see that $B^{(\ga)}$ also encode the fusion algebra
\begin{align*}
\bpm
B^{(0)} |0\> & B^{(0)} |1\> & B^{(0)} |2\> \\
B^{(1)} |0\> & B^{(1)} |1\> & B^{(1)} |2\> \\
B^{(2)} |0\> & B^{(2)} |1\> & B^{(2)} |2\> \\
\epm
=
\bpm
|0\> &  |1\> & |2\> \\
|1\> &  |0\> & |2\> \\
|2\> &  |2\> & |0\>+|1\> \\
\epm ,
\end{align*}
where $|\ga\>$, $\ga=0,1,2$  are the degenerate ground states.

For the solution $S_2$ we have
\begin{align*}
 (a^{(\ga')}_\ga)=
\bpm
1&1& -\sqrt 2 \\
1&1&\sqrt 2 \\
1&-1&0\\
\epm
\end{align*}
and
\begin{align*}
 A^{(0)}&=
\bpm
1&0&0\\
0&1&0\\
0&0&1\\
\epm,\ \ \ \ \
 A^{(1)}=
\bpm
1&0&0\\
0&1&0\\
0&0&-1\\
\epm,
\nonumber\\
 A^{(2)} &=
\bpm
-\sqrt 2&0&0\\
0&\sqrt 2&0\\
0&0&0\\
\epm  .
\end{align*}
$B^{(\ga)}$ remain the same.

The first row of the $S$-matrix are called quantum dimensions of
the quasi-particles.  For a unitary topological quantum field
theory, all quantum dimensions must be positive real numbers, and
moreover $\geq 1$. Therefore, the second solution does not give
rise to a unitary topological field theory.  Based on this reason,
we exclude the $S_2$ solution.

The $\nu=1$ bosonic $Z_2$ parafermion state $\Phi_{\frac22;Z_2}$ has three
degenerate ground states on a torus.  In the thin torus limit, the three ground
states are described by the occupation distributions $20202020\cdots$,
$02020202\cdots$, and $11111111\cdots$.

The $\nu=1/2$ fermionic $Z_2$ parafermion state $\Phi_{\frac22;Z_2}\prod
(z_i-z_j)$ has six degenerate ground states on a torus.  In the thin torus
limit, the six ground states are also described by the occupation
distributions.  Those occupation distributions can be obtained from that of
$\Phi_{\frac22;Z_2}$ state given above.  We note that multiplying the factor
$\prod (z_i-z_j)$ increases the space between every neighboring particles in a
distribution by one. For example, it changes $11$ to $101$, $101$ to $1001$,
$2$ to $11$, $3$ to $111$, \etc.  It changes the bosonic distribution
$202020\cdots$ to a fermionc distribution $110011001100\cdots$, and changes
the distribution $111111\cdots$ to $101010101010\cdots$.  Including the
translated distributions of $110011001100\cdots$ and $101010101010\cdots$, we
find that the fermionic $Z_2$ parafermion state $\Phi_{\frac22;Z_2}\prod
(z_i-z_j)$ has six degenerate ground states described by the distributions
\begin{align*}
& 110011001100\cdots,
\nonumber\\
& 011001100110\cdots,
\nonumber\\
& 001100110011\cdots,
\nonumber\\
& 100110011001\cdots,
\nonumber\\
& 101010101010\cdots,
\nonumber\\
& 010101010101\cdots.
\end{align*}
Note that a unit cell contains $m=4$ (or 2) orbitals.

\subsection{$Z_3$ parafermion state}

The $\nu=3/2$ bosonic $Z_3$ parafermion state
has a pattern of zeros described by
\begin{align*}
\Phi_{\frac{3}{2};Z_3}:\  (m;S_2,\cdots,S_n) &=(2;0,0),
\nonumber\\
(\frac{m}{n}; h^\text{sc}_1,\cdots,h^\text{sc}_n) &=(\frac{2}{3}; \frac23,\frac23,0),
\nonumber\\
(n_0,\cdots,n_{m-1})&=(3,0)
\end{align*}
Solving \eq{SqCond}, we find that
there are four types of quasiparticles:
\begin{align*}
 \begin{tabular}{|c|c|c|}
\hline
$n_{\ga;l}$ & $Q_\ga$ & $S^\text{osp}_\ga$ \\
\hline
       30  &  0  &  0   \\
       03  &  3/2  &  3/4   \\
       12  &  1  &  1/2   \\
       21  &  1/2  &  1/4  \\
\hline
 \end{tabular}
\end{align*}
Thus in the $|\ga\>$ basis, $T_1$, $T_2$ and $T$ are given by
\begin{align*}
 T_1 &=
\bpm
1 & 0 & 0 & 0 \\
0& -1 & 0 & 0 \\
0&  0 & 1 & 0 \\
0&  0 & 0  & -1 \\
\epm,\ \ \ \
 T_2=
\bpm
0 & 1 & 0 & 0 \\
1 & 0 & 0 & 0 \\
0 & 0 & 0 & 1 \\
0 & 0 & 1 & 0 \\
\epm,
\nonumber\\
 T &=
\bpm
1&  0     & 0     & 0 \\
0&  t_1 & 0     & 0 \\
0&  0     & t    & 0 \\
0&  0     & 0     & t_1t_2 \\
\epm  .
\end{align*}
Since $l_{max}=0$, from \eq{nbarganga}, we find that
$\overline{30}=30$,
$\overline{03}=03$,
$\overline{21}=21$, and
$\overline{12}=12$.
Thus $ C=1 $.

From $TT_2=\e^{\i \pi n/m} T_2T_1 T$ (see \eqn{TST1T2}) we find that $t_1=-\i$
and $t_2=t$.  Thus
\begin{equation*}
 T =
\bpm
1&  0     & 0     & 0 \\
0&  -\i & 0     & 0 \\
0&  0     & t    & 0 \\
0&  0     & 0     & -\i t \\
\epm  .
\end{equation*}
We can rewrite $T_1$, $T_2$, and $T$ in direct product form
\begin{align*}
 T_1&=\si^3\otimes\si^0,  & T_2&=\si^1\otimes\si^0,
\nonumber\\
 T &=
\bpm
1&0\\
0&-\imth
\epm
\otimes
\bpm
1&0\\
0&t
\epm
.
\end{align*}
Using $ST_1=T_2S$, $ST_2=T_1S$, and $S^2=1$,
we find that $S$ must have the following form
\begin{align*}
S&=
\frac{1}{\sqrt 2}
\bpm
1&1\\
1&-1
\epm
\otimes
\bpm
\cos(\th)&\sin(\th)\\
\sin(\th)&-\cos(\th)
\epm
.
\end{align*}
Note that the above $S$ is already symmetric.
Thus we can regard the above $S$ as the $S^\text{TC}$.

The direct product form of $T_1$, $T_2$, $T$, and $S$ suggests
that the $U(1)$ charge part and the non-Abelian part separate.\cite{DFL} So
let us concentrate on the non-Abelian part
\begin{equation*}
 \t T=
\bpm
1&0\\
0&t
\epm ,\ \ \ \ \
 \t S
=
\bpm
\cos(\th)& \sin(\th)\\
\sin(\th)&-\cos(\th)
\epm  .
\end{equation*}
From $\t S$, we can calculate $\t A^{(\ga)}$ and $\t B^{(\ga)}$ [see \eqn{Sba}
and \eqn{TSAB}]. All the matrix elements $\t B^{(\ga)}$ must be non-negative
integers and $(\t S\t T)^3\propto 1$.

One way to satisfy those conditions is to let
\begin{equation*}
 \tan(\th)=\frac{1+\sqrt 5}{2} \equiv \vphi.
\end{equation*}
In this case
\begin{equation}
\label{tTtS}
 \t T=
\bpm
1&0\\
0&\e^{\pm \i 4 \pi/5}
\epm ,\ \ \ \ \
 \t S
=\frac{1}{\sqrt{1+\vphi^2}}
\bpm
1     & \vphi \\
\vphi & -1
\epm  .
\end{equation}
and
\begin{align*}
\t A^{(0)}&=1,
&
\t A^{(1)}&=
\bpm
\vphi & 0\\
0 & 1-\vphi
\epm  ,
\nonumber\\
\t B^{(0)}&=1,
&
\t B^{(1)}&=
\bpm
0 & 1\\
1 & 1
\epm  .
\end{align*}

Those are the only valid solutions, which are realized by the Fibonacci
anyons.  The solutions with $d_1<0$ are excluded as reasoned at the end of the
last section, and one of the them is the non-unitary Yang-Lee model.

Taking $\ga_1=\ga_2=1$ in \eqn{Shhh}, we find that
\begin{align*}
 -1=\Big(
\e^{-\i 4\pi \t h_1}
+\e^{-\i 2\pi \t h_1} \vphi
\Big)
\end{align*}
or
\begin{equation*}
 \cos(2\pi \t h_1)=-\vphi/2 ,
\end{equation*}
which gives us
\begin{equation*}
 \t h_1 = \pm 2/5 \text{ mod } 1.
\end{equation*}
Since $\t h_0=0$, we see that $\e^{\i 2\pi \t h_\ga}$ are the
eigenvalues of the $\t T$ operator (see \eqn{tTtS}). Let us denote
the eigenvalues of the $T$ operator as $\e^{\i 2\pi h^T_\ga}$. We
find, for the two choices of $\t h^T_1 = \pm 2/5$ (see \eqn{tTtS}),
\begin{equation*}
 \begin{tabular}{|c|c|c|}
\hline
$n_{\ga;l}$ & $h^T_\ga$ & $h^T_\ga$ \\
\hline
       30  &  0  &  0   \\
       03  &  3/4  &  3/4   \\
       12  &  2/5  &  3/5   \\
       21  &  3/20  &  7/20  \\
\hline
 \end{tabular}
\end{equation*}
According to CFT (see appendix \ref{cftsec}), the scaling dimension of the
quasiparticle operators in the $Z_3$ parafermion FQH state is
$h_\ga=\frac{Q_\ga^2}{3}+ h^\text{sc}_\ga$.  For the quasiparticles 30 and 03,
$h^\text{sc}_\ga=0$ since those are Abelian quasiparticles.  For the
quasiparticles 12 and 21, $h^\text{sc}_\ga=1/15$.  Thus $h_\ga = 0$, 3/4, 2/5,
and 3/20 for the quasiparticles 03, 30,12, and 21 respectively, which exactly
agree with $h^T_\ga$ for the case of $\t h^T_1 =  2/5$.  This example
demonstrates a way to calculate\cite{RSW} quasiparticle scaling dimensions
from the pattern of zeros.

The $\nu=3/2$ bosonic $Z_3$ parafermion state $\Phi_{\frac32;Z_3}$ has four
degenerate ground states on a torus, described by the occupation distributions
\begin{align*}
 3030303030\cdots,
\nonumber\\
 0303030303\cdots,
\nonumber\\
 2121212121\cdots,
\nonumber\\
 1212121212\cdots,
\end{align*}
The $\nu=3/5$ fermionc $Z_3$ parafermion state $\Phi_{\frac32;Z_3}\prod
(z_i-z_j)$ has 10
degenerate ground states on a torus, described by the occupation distributions
\begin{align*}
 1110011100111001110011100\cdots,
\nonumber\\
 0111001110011100111001110\cdots,
\nonumber\\
 0011100111001110011100111\cdots,
\nonumber\\
 1001110011100111001110011\cdots,
\nonumber\\
 1100111001110011100111001\cdots,
\nonumber\\
 1101011010110101101011010\cdots,
\nonumber\\
 0110101101011010110101101\cdots,
\nonumber\\
 1011010110101101011010110\cdots,
\nonumber\\
 0101101011010110101101011\cdots,
\nonumber\\
 1010110101101011010110101\cdots.
\end{align*}
Note that a unit cell contains $m=5$ orbitals.

\section{Conclusions}

Through string-net wave functions\cite{FNS0428,LWstrnet}, one can show that
non-chiral topological orders can be naturally described and classified by
tensor category theory.  This raises a question: how to describe and classify
the chiral topological order in FQH states? The results in \Ref{WWsymm} suggest
that the pattern of zeros may provide a way to characterize and classify
chiral topological orders in FQH states.  In this paper, we see that many
topological properties of chiral topological orders can be calculated from the
data $\{S_a\}$ that describe the pattern of zeros.  In particular, through the
algebra of tunneling operators, we see a close connection to tensor category
theory.  The pattern of zeros provides a link from electron wave functions (the
symmetric polynomials) to tensor category theory and the corresponding chiral
topological orders.

\begin{acknowledgments}
We would like to thank M. Freedman and N. Read for helpful discussions.
This research is partially supported by NSF Grant No.  DMR-0706078 (XGW)
and by NSF Grant No. DMS-034772 (ZHW).
\end{acknowledgments}

\appendix

\section*{Appendix}

\subsection{The calculation of orbital spin}

\label{calosp}

Let
\begin{align}
\label{FlJ}
 F_{L,J}(\al)&=\sum_{l=0} (l-J) \e^{-\al^2 l^2} \del_{l\%m - L}
\nonumber\\
&=\sum_{k=0}  (km+L-J) \e^{-\al^2 (km+L)^2} .
\end{align}
We find that, for a canonical occupation distribution \eq{ngalcan},
we can rewrite \eq{thBqR} as
\begin{equation}
\label{thBqF}
\th^\ga_B=\Om \sum_{l=0}^{m-1} (n_{\ga;l}-n_l) F_l(\al) .
\end{equation}
To evaluate $F_{L,J}(\al)$, we will use the Euler-Maclaurin formula:
\begin{align*}
&\ \ \ \  \frac{f(0)+f(k_{max})}{2}+\sum_{k=1}^{k_{max}} f(k)-
\int_0^{k_{max}} f(x) \dd x
\nonumber\\
&= \sum_{k=1}^p \frac{B_{k+1}}{(k+1)!}
\Big(f^{(k)}(k_{max})-f^{(k)}(0) \Big) + R,
\end{align*}
where $B_k$ are Bernoulli numbers
$(B_0,B_1,B_2,B_3,\cdots)=(1,-\frac12,\frac{1}{6},0,-\frac{1}{30},\cdots)$,
and
\begin{equation*}
 |R|\leq \frac{2}{(2\pi)^{2p}} \int_0^n |f^{(2p+1)}(x)| \dd x .
\end{equation*}
For our case $f(x)=(L+mx-J)\e^{-\al^2(mx+L)^2}$.  If we choose $\al\sim 1/J$
and $p=1$, in large $J$ limit, we find $R\sim 1/J$ in large $J$ limit.
Therefore
\begin{align*}
&\ \ \ \ F_{L,J}(\al)
\nonumber\\
 &=
\int_0^\infty  (L+mx-J)\e^{-\al^2(mx+L)^2} \dd x
\nonumber\\
&\ \ \ \ \ \ \ \ \ \
+\frac 12 (L-J) -\frac{m}{12}+O(1/J)
\nonumber\\
&=
\int_0^\infty  \frac{L+\frac{x}{\al} -J}{m\al}\e^{-(x+\al L)^2} \dd x
\nonumber\\
&\ \ \ \ \ \ \ \ \ \
+\frac 12 (L-J) -\frac{m}{12}+O(1/J)
\nonumber\\
&=
\int_0^\infty
\frac{L+\frac{x}{\al}-J}{m\al}
(1-2\al L x-\al^2L^2+ 2\al^2L^2 x^2)\e^{-x^2} \dd x
\nonumber\\
&\ \ \ \ \ \ \ \ \ \
+\frac 12 (L-J) -\frac{m}{12}+O(1/J)
\nonumber\\
&=
-\frac{\al J \sqrt\pi -1}{2\al^2 m}
+\frac{ J }{ m} L
-\frac{L^2}{2m}
+\frac {L-J}{2} -\frac{m}{12}+O(1/J)
.
\end{align*}
Since $\sum_{l=0}^{m-1} (n_l - n_{\ga;l}) = 0$ for a canonical
occupation distribution, the terms that do not depend on $L$ do not
contribute to the Berry's phase.  Thus we have
\begin{align}
\label{thBql0}
\th^\ga_B=\Om \sum_{l=0}^{m-1} (n_{\ga;l}-n_l)
\Big(\frac{J}{m} l +\frac{l}{2}-\frac{l^2}{2m}\Big)  .
\end{align}
Compare with \eq{Qqnlq} and \eq{thBQq}, we find that the $\frac{J}{m} l$ term
exactly reproduces the quasiparticle charge.
The terms that do not depend on $J$ give us the orbital spin \eq{ospnl}.

Since $n_l$ is a periodic function of $l$ with a period $m$, we may also view
$n_l$ as a periodic function with a period $km$.  If we view $n_l$ as a
periodic function with a period $km$, the  orbital spin $S_\ga^\text{osp}$ will be
given by
\begin{align*}
 S_\ga^\text{osp}&=
\sum_{l=0}^{km-1} (n_{\ga;l} - n_l) \Big(\frac{l}{2}-\frac{l^2}{2km}\Big)
\nonumber\\
&=
\sum_{j=0}^{k-1} \sum_{l=0}^{m-1}
(n_{\ga;l} - n_l) \Big(\frac{jm+l}{2}-\frac{(jm+l)^2}{2km}\Big)
\nonumber\\
&=
\sum_{j=0}^{k-1} \sum_{l=0}^{m-1}
(n_{\ga;l} - n_l) \Big( \frac{l}{2} -\frac{j l}{k} - \frac{l^2}{2km} \Big)
\nonumber\\
&=
\sum_{l=0}^{m-1}
(n_{\ga;l} - n_l) \Big( \frac{kl}{2} -\frac{k(k-1) l}{2k} - \frac{l^2}{2m} \Big)
\nonumber\\
&=
\sum_{l=0}^{m-1}
(n_{\ga;l} - n_l) \Big( \frac{l}{2} - \frac{l^2}{2m} \Big) ,
\end{align*}
which is identical to the previous result \eq{ospnl}.  Therefore,
the formal calculation of $S_\ga^\text{osp}$ using $\e^{-\al^2 l^2}$
regulator (see \eq{thBqR}) produces a sensible result.

\subsection{Orbital spin of Abelian quasiparticles}
\label{ospAb}

\begin{figure}[t]
\centerline{
\includegraphics[scale=0.55]{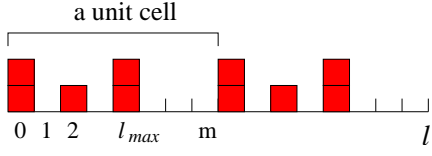}
}
\caption{
The graphic representation of an occupation distribution
$(n_0,\cdots,n_{10})=(2,0,1,0,2,0,0,0)$ for the
$\Phi_{\frac{5}{8};Z_5^{(2)}}$ state
with $n=5$ and $m=8$.
$l_{max}=4$ for such a distribution.
}
\label{nldist}
\end{figure}

If a quasiparticle is described by an occupation distribution $n_{\ga;l}$ that
can be obtained by shifting the occupation distribution $n_l$ for the ground
state,
then the orbital spin of such a quasiparticle can be calculated reliably
without using the formal unreliable approach described above. Such kind of
quasiparticles can be created by threading magnetic flux lines through the FQH
liquid and correspond to Abelian quasiparticles.

The occupation distribution $n_l$ for the ground state has some properties
that will be important for the following discussion.  In addition to the
periodic property $n_{l+m}=n_l$, $n_l$ also have a symmetric property
\begin{equation}
\label{nlsym}
 n_l=n_{l_{max}-l},\ \ \ \ \ 0 \leq l \leq l_{max}.
\end{equation}
according to numerical experiments, where $l_{max}=S_n-S_{n-1}$ is the largest
$l$ in the first unit cell $0\leq l < m$ such that $n_{l_{max}}>0$ (see Fig.
\ref{nldist}).  (\eq{nlsym} implies that $n_0 > 0$).

\begin{figure}[t]
\centerline{
\includegraphics[scale=0.55]{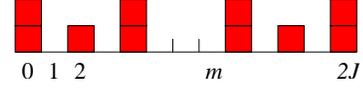}
}
\caption{
If $N_\phi=2J=l_{max}+N_c m$, the symmetric polynomial $\Phi$
of $N=N_c n$ variables can be put on the sphere without any defects
(or quasiparticles).
}
\label{nlsph}
\end{figure}

\begin{figure}[t]
\centerline{
\includegraphics[scale=0.55]{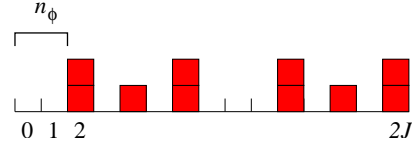}
}
\caption{
If $N_\phi=2J=l_{max}+(N_c-1) m+n_\phi$, the above occupation distribution
$n_{\ga;l}$ (obtained by shifting the ground state distribution in Fig.
\ref{nlsph}) describes a single quasiparticle on the sphere located at the
south pole ($l=0$).  There is no quasiparticle at the north pole ($l=2J$).
}
\label{nlsphq}
\end{figure}

On a sphere with $N_\phi=2J$ flux quanta, there are $N_\phi+1$
orbitals labeled by $l=0,1,\cdots,2J$ (see Fig. \ref{mzl}).  Those
orbitals form a angular momentum $J$ representation of the $O(3)$
rotation.  The $J_z$ quantum numbers of those orbitals are given by
$m_z=l-J$.  If the flux through the sphere is such that
$N_\phi=2J=l_{max}+(N_c-1) m$ for an integer $N_c$, then the
occupation $n_l$ can fit into the $2J+1$ orbital in such a way that
there is no non-trivial quasiparticle at the north and south poles
(see Fig. \eq{nlsph}).  Such a state has $N=N_c n$ particles. Such
an $N_c n$-particle state can fill the sphere without any defect and
form a $J^\text{tot}=0$ state.  Note that $l_{max}=l_n=S_n-S_{n-1}$
and $S_{n-1}=\frac{n-2}{n}S_n$ (see \Ref{WWsymm}), thus $l_{max} =
\frac{2}{n}S_n$.  So an $N_c n$-particle state can fill the sphere
without any defect if $N_\phi=2J=\frac{2}{n}S_n+(N_c-1) m$ which is
exactly the condition obtained in \Ref{WWsymm}.

Let us create a quasiparticle by threading $n_\phi$ flux lines
through the south pole.  The total flux quanta becomes
$N_\phi=2J=\frac{2}{n}S_n+(N_c-1) m+n_\phi$ and the occupation
distribution $n_{\ga;l}$ for the created quasiparticle is obtained
from $n_l$ by shifting the distribution by $n_\phi$ (see Fig.
\ref{nlsphq}).  The occupation distribution is identical to that of
the ground state distribution in Fig. \ref{nlsph} near the north
pole ($l=2J$). Thus the distribution describe a state that has no
quasiparticle near the north pole.  However, the occupation
distribution is different from the ground state distribution near
the south pole ($l=0$).  Therefore, the distribution describes a
state with a quasiparticle near the south pole.

The total $J_z$ of the above quasiparticle state is given by
\begin{align*}
 J^\text{tot}_z &= \sum_{l=0}^{2J} n_{\ga;l} m_z = \sum_{l=0}^{2J} n_{\ga;l} (l-J)
=\frac{n_\phi}{2} N
\nonumber\\
&=
\frac{n_\phi n}{m}
\Big(
J+\frac{m-l_{max}}{2}-\frac{n_\phi}{2}
\Big)
\nonumber\\
&=
J Q_\ga + \frac{m-l_{max}}{2} Q_\ga - \frac{m}{2n} Q_\ga^2
\end{align*}
where
\begin{equation}
\label{Qqnphi}
Q_\ga=\frac{n_\phi n}{m}
\end{equation}
is the quasiparticle charge.  If we move such quasiparticle along a loop that
spans a solid angle $\Om$, the induced Berry's phase $\th_B^\ga$ will be
\begin{equation*}
 \frac{\th_B^\ga}{\Om}=J^\text{tot}_z=
J Q_\ga + \frac{m-l_{max}}{2} Q_\ga - \frac{m}{2n} Q_\ga^2  .
\end{equation*}
Compare with \eq{thBSosp}, we find the orbital spin of the quasiparticle to be
\begin{equation}
\label{ospAbl}
 S_\ga^\text{osp}= \frac{m-l_{max}}{2} Q_\ga - \frac{m}{2n} Q_\ga^2  .
\end{equation}
Let us compare \eq{ospAbl} with \eq{ospnl}.  For the Abelian quasiparticle,
its occupation distribution $n_{\ga;l}$ has a form
\begin{equation*}
 n_{\ga;l}=n_{l-n_\phi} .
\end{equation*}
In this case, \eq{ospnl} becomes
\begin{align*}
 S_\ga^\text{osp}&=\sum_{l=0}^{m-1} n_l
\Big(
\frac{l+n_\phi}{2}-\frac{(l+n_\phi)^2}{2m} -\frac{l}{2}+\frac{l^2}{2m}
\Big)
\nonumber\\
&=
\frac{ n n_\phi}{2} -\frac{n}{2m} n_\phi^2
-\frac{n_\phi}{m} \sum_{l=0}^{m-1} l n_l
.
\end{align*}
Using \eq{nlsym}, we find that the above expression agrees with \eq{ospAbl}.
This confirms the validity of \eq{ospnl} and \eq{ospla} for the case of
Abelian quasiparticles.  On the other hand, the validity of \eq{ospnl} or
\eq{ospla} for the case of non-Abelian quasiparticle is yet to be confirmed by
a more rigorous calculation.

\subsection{Translations of orbitals on torus}
\label{T1T2act}

To show $T_2\phi^{(l)}= \phi^{(l+1)}$, we note that
\begin{align*}
&\ \ \ \ T_2\phi^{(l)}(X^1,X^2)
\nonumber\\
&=
\e^{\imth 2\pi X^1}
f^{(l)}\Big(X^1+\tau (X^2 +\frac{1}{N_\phi})\Big)
\e^{\imth \pi N_\phi \tau(X^2+\frac{1}{N_\phi})^2} .
\end{align*}
Since
\begin{align*}
&\ \  f^{(l)}(z+\frac{\tau}{N_\phi})
=\sum_{k} \e^{ \imth\frac{\pi \tau}{N_\phi} (N_\phi k + l)^2
+ \imth 2  \pi (N_\phi k +l) (z+\frac{\tau}{N_\phi} ) }
\nonumber\\
&=
\e^{- \imth\frac{\pi \tau}{N_\phi}}
\sum_{k} \e^{ \imth\frac{\pi \tau}{N_\phi} (N_\phi k + l+1)^2
+ \imth 2  \pi (N_\phi k +l) z }
\nonumber\\
&=
\e^{- \imth\frac{\pi \tau}{N_\phi}}
\e^{-\imth 2\pi z}
f^{(l+1)}(z)
,
\end{align*}
we have
\begin{align*}
&\ \ \ \ T_2\phi^{(l)}(X^1,X^2)
\nonumber\\
&=
\e^{\imth 2\pi X^1
- \imth\frac{\pi \tau}{N_\phi}
-\imth 2\pi z}
f^{(l+1)}(X^1+\tau X^2)
\e^{\imth \pi N_\phi \tau(X^2+\frac{1}{N_\phi})^2}
\nonumber\\
&= \phi^{(l+1)}(X^1,X^2)
.
\end{align*}

\subsection{Non-Abelian Berry's phase and modular transformation}
\label{nabmodu}

The wave functions $\Phi_{\{n_{\ga;l+l_s}\}}$ form a basis of the
degenerate ground states.  As we change the mass matrix or $\tau$,
we obtain a family of basis parameterized by $\tau$.  The family of
basis can give rise to non-Abelian Berry's phase\cite{WZ8411} which
contain a lot information on topological order in the FQH state. In
the following, we will discuss such a non-Abelian  Berry's phase in
a general setting. We will use $\ga$ to label the degenerate ground
states.

To find the non-Abelian Berry's phase, let us first define parallel
transportation of a basis.  Consider a path $\tau(s)$ that deform
the inverse-mass-matrix $g(\tau_1)$ to $g(\tau_2)$: $\tau_1=\tau(0)$
and $\tau_2=\tau(1)$.  Assume that for each inverse-mass-matrix
$g[\tau(s)]$, the many-electron Hamiltonian on torus $ (X^1,X^2)\sim
(X^1+1,X^2)\sim (X^1,X^2+1) $ has $N_q$-fold degenerate ground
states $|\ga;s\>$, $\ga=0,1,\cdots,N_q-1$, and a finite energy gap
for excitations above the ground states.  We can always choose a
basis $|\ga;s\>$ for the ground states such that the basis for
different $s$ satisfy
\begin{equation*}
\<\ga';s|\frac{\dd}{\dd s} |\ga;s\> =0.
\end{equation*}
Such a choice of basis $|\ga;s\>$ defines a parallel transportation
from the bases for inverse-mass-matrix $g(\tau_1)$ to that for
inverse-mass-matrix $g(\tau_2)$ along the path $\tau(s)$.

In general, the parallel transportation is path dependent.  If we
choose another path $\tau'(s)$ that connect $\tau_1$ and $\tau_2$,
the parallel transportation of the same basis for
inverse-mass-matrix  $g(\tau_1)$, $|\ga;s=0\>=|\ga;s=0\>'$, may
result in a different basis for inverse-mass-matrix $g(\tau_2)$,
$|\ga;s=1\>\neq |\ga;s=1\>'$.  The different basis are related by an
invertible transformation. Such a path dependent invertible
transformation is the non-Abelian Berry's phase.\cite{WZ8411}

However, for the degenerate ground states of a topologically ordered
state (including a FQH state), the  parallel transportation has a
special property that, up to a total phase, it is path independent
(in the thermal dynamical limit).  The parallel transportations
along different paths connecting $\tau_1$ and $\tau_2$ will change a
basis for inverse-mass-matrix $g(\tau_1)$ to the same basis for
inverse-mass-matrix $\tau(\tau_2)$ up to an overall phase:
$|\ga;s=1\>=\e^{\imth\phi} |\ga;s=1\>'$.
In particular, if we deform an inverse-mass-matrix through a loop into itself
(\ie $\tau(0)=\tau(1)$), the basis $|\ga;0\>$ will parallel transport into
$|\ga;1\>=\e^{\imth\vphi}|\ga;0\>$.  Thus, non-Abelian Berry's phases for the
degenerate states of a topologically ordered state are only path-dependent
Abelian phases $\e^{\imth\vphi}$ which do not contain much information of
topological order.

However, there is a class of special pathes which give rise to non-trivial
non-Abelian Berry's phases.  First we note that the torus $(X^1,X^2)\sim
(X^1+1,X^2) \sim (X^1,X^2+1)$ can be parameterized by another set of
coordinates
\begin{align}
\label{XpX}
&
\bpm X^{\prime 1}\\ X^{\prime 2}\\ \epm
=
\bpm
d&-b\\
-c&a\\
\epm
 \bpm X^1\\ X^2\\ \epm,
\ \
\bpm X^1\\ X^2\\ \epm
=
\bpm
a&b\\
c&d\\
\epm
 \bpm X^{\prime 1}\\ X^{\prime 2}\\ \epm,
\end{align}
where $a,b,c,d \in Z$,  $ad-bc=1 $. The above can be rewritten in
vector form
\begin{align}
\label{XpXV}
&
\v X' = M^{-1}\v X,
\ \ \ \ \ \ \
\v X = M\v X',
\ \ \ \
M\in SL(2,Z) .
\end{align}
$(X^{\prime 1},X^{\prime 2})$ has the same periodicity condition
$(X^{\prime 1},X^{\prime 2})\sim (X^{\prime 1}+1,X^{\prime 2}) \sim
(X^{\prime 1},X^{\prime 2}+1)$ as that for $(X^1,X^2)$.  We note
that
\begin{equation*}
  \bpm \prt_{X^1}\\ \prt_{X^2}\\ \epm
=
\bpm
d&-c\\
-b&a\\
\epm
 \bpm \prt_{X^{\prime 1}}\\ \prt_{X^{\prime 2}}\\ \epm
,\ \ \ \
  \bpm \prt_{X^{\prime 1}}\\ \prt_{X^{\prime 2}}\\ \epm
=
\bpm
a&c\\
b&d\\
\epm
 \bpm \prt_{X^1}\\ \prt_{X^2}\\ \epm
.
\end{equation*}
The inverse-mass-matrix in the $(X^1,X^2)$ coordinate, $g(\tau)$,
is changed to
\begin{equation*}
 g'=
\bpm
d&-b\\
-c&a\\
\epm
g(\tau)
\bpm
d&-c\\
-b&a\\
\epm
\end{equation*}
in the $(X^{\prime 1},X^{\prime 2})$ coordinate.
From \eqn{mtau}, we find that
\begin{equation}
 g'=
\bpm
\tau_y'+\frac{{\tau_x'}^2}{\tau_y'} & -\frac{\tau_x'}{\tau_y'}\\
-\frac{\tau_x'}{\tau_y'} & \frac{1}{\tau_y'}
\epm = g(\tau'),
\end{equation}
with
\begin{equation}
\label{tauptau}
 \tau'=\frac{b+d\tau}{a+c\tau} .
\end{equation}
The above transformation $\tau\to \tau'$ is the modular transformation.  We
see that if $\tau$ and $\tau'$ are related by the modular transformation, then
two inverse-mass-matrices $g(\tau)$ and $g(\tau')$ will actually describe the
same system (upto a coordinate transformation).

\begin{figure}[t]
\centerline{
\includegraphics[scale=0.41]{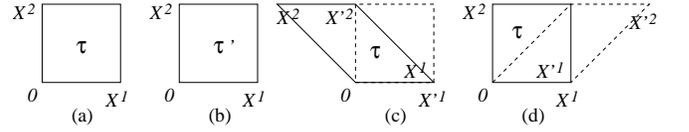}
} \caption{ (a): A system described by an inverse-mass-matrix
$g(\tau)$. (b): A system described by an inverse-mass-matrix
$g(\tau')$ where $\tau$ and $\tau'$ are related by a modular
transformation $M_T$. (c): The system in (b) is the same as the
system in (a) if we make a change in coordinates. (d): The system in
(c) is drawn differently. } \label{modutran}
\end{figure}

Let us assume that the path $\tau(s)$ connects two $\tau$'s related
by a modular transformation $M=\bpm a&b\\ c&d \epm$:
$\tau(1)=\frac{b+d\tau(0)}{a+c\tau(0)}$.  We will denote
$\tau(0)=\tau$ and $\tau(1)=\tau'$.  The parallel transportation of
the basis $|\ga;\tau\>$ for inverse-mass-matrix $g[\tau(0)]$ gives
us a basis $|\ga;\tau'\>$ for inverse-mass-matrix $g[\tau(1)]$.
Since $\tau=\tau(0)$ and $\tau'=\tau(1)$ are related by a modular
transformation, $g[\tau(0)]$ and $g[\tau(1)]$ actually describe the
same system. The two basis $|\ga;\tau\>$ and $|\ga;\tau'\>$ are
actually two basis of same space of the degenerate ground states.
Thus there is an invertible matrix that relate the two basis
\begin{align}
\label{alUbt}
 |\ga;\tau'\> &=U(M) |\ga;\tau\>
\\
U_{\ga\ga'}(M)
&
=\<\ga;\tau|\ga';\tau'\>
=\<\ga;\tau|U(M)|\ga';\tau\>
.
\nonumber
\end{align}
Such an invertible matrix is the non-Abelian Berry's phase for the path $\tau(s)$.
Except for its overall phase (which is path dependent), the invertible matrix $U$
is a function of the modular transformation $M$.  In fact, the invertible matrix
$U$ form a projective representation of the modular transformation.  The
projective representation of the modular transformation contains a lot of
information of the underlying topological order.

Let us examine $\<\ga;\tau|\ga';\tau\>$ in \eqn{alUbt} more
carefully.  Let $\Phi_{\ga}[\{\v X_i\}|\tau]$ be ground state wave
functions for inverse-mass-matrix $g[\tau]$, and $\Phi_{\ga}[\{\v
X_i\}|\tau']$ be ground state wave functions for inverse-mass-matrix
$g[\tau']$.  Here $\v X_i=(X^1_i,X^2_i)$ are the coordinates of the
$i^\text{th}$ electron. Since $\tau$ and $\tau'$ are related by a
modular transformation, $\Phi_{\ga}[\{\v X_i\}|\tau]$ and
$\Phi_{\ga}[\{\v X_i\}|\tau']$ are ground state wave function of the
same system.  However, we cannot directly compare $\Phi_{\ga}[\{\v
X_i\}|\tau]$ and $\Phi_{\ga}[\{\v X_i\}|\tau']$ and calculate the
inner product between the two wave functions as
\begin{equation*}
\Big(\Phi_{\ga}[\{\v X_i\}|\tau(0)],\ \Phi_{\ga'}[\{\v X_i\}|\tau'] \Big).
\end{equation*}
The wave function $\Phi_{\ga}[\{\v X_i\}|\tau']$ for inverse-mass-matrix
$g[\tau']$ can be viewed as the ground state wave function for
inverse-mass-matrix $g[\tau]$ only after a coordinate transformation (see Fig.
\ref{modutran}).  Let us
rename $\v X$ to $\v X'$ and rewrite $\Phi_{\ga}[\{\v X_i\}|\tau']$ as
$\Phi_{\ga}(\{\v X^{\prime}_i\}|\tau')$.  Since the coordinate transformation
\eq{XpXV} change $\tau$ to $\tau'$, we see that we should really compare
$\Phi_{\ga}(\{\v X^{\prime}_i\}|\tau')=\Phi_{\ga}(\{M^{-1}\v X_i\}|\tau')$
with $\Phi_{\ga}(\{\v X_i\}|\tau)$.
But even $\Phi_{\ga}(\{\v X_i\}|\tau)$ and
$\Phi_{\ga}(\{M^{-1}\v X_i\}|\tau')$ cannot be directly compared.  This
is because the coordinate transformation \eq{XpX} changes the gauge potential
\eq{AXAY} to another gauge equivalent form.  We need to perform a $U(1)$ gauge
transformation $U_G(M)$ to transform the changed gauge potential back to its
original form \eqn{AXAY}. So only
$\Phi_{\ga}[\{\v X_i\}|\tau]$ and
$U_G\Phi_{\ga}[\{M^{-1}\v X_i\}|\tau']$ can be directly compared.
Therefore, we have
\begin{align}
\label{PsiUPsi}
&\ \ \ \ U_{\ga\ga'}(M)
\\
&=
\Big( \Phi_{\ga}(\{\v X_i\}|\tau),
 U_G(M)\Phi_{\ga'}(\{M^{-1}\v X_i\}|\tau')
\Big)
\nonumber
\end{align}
which is \eqn{alUbt} in wave function form.
Note that $\tau'=\frac{M_{12}+M_{22}\tau}{M_{11}+M_{21}\tau}$.

Let us calculate the gauge transformation $U_G(M)$.
We note that
$\Phi_{\ga}(\{\v X'_i\}|\tau')$ is the ground state of
\begin{equation*}
 H'=-\sum_k
\frac{1}{2}\sum_{i,j=1,2}
\Big(\frac{\partial}{\partial X^{\prime i}_k} - \imth A'_i \Big)
g'_{ij}
\Big(\frac{\partial}{\partial X^{\prime j}_k} - \imth A'_j \Big)
\end{equation*}
where $k=1,\cdots,N$ labels the different electrons.
In terms of $X^i$ (see \eqn{XpX}), $H'$ has a form
\begin{equation*}
 H'=-\sum_k
\frac{1}{2}\sum_{i,j=1,2}
\Big(\frac{\partial}{\partial X^{i}_k} - \imth \t A_i \Big)
g_{ij}
\Big(\frac{\partial}{\partial X^{j}_k} - \imth \t A_j \Big)
\end{equation*}
where
\begin{equation*}
  \bpm \t A_1\\ \t A_2\\ \epm
=
\bpm
d&-c\\
-b&a\\
\epm
 \bpm A'_1\\ A'_2\\ \epm
,\ \ \ \
  \bpm A'_1 \\ A'_2\\ \epm
=
\bpm
a&c\\
b&d\\
\epm
 \bpm \t A_1 \\ \t A_2\\ \epm
.
\end{equation*}
Since
$(A'_1,A'_2)=(-2\pi N_\phi X^{\prime 2},0)$,
We find that
\begin{align*}
&\ \ \ \ (\t A_1,\t A_2)=
( -2\pi N_\phi X^{\prime 2} d,\ 2\pi N_\phi X^{\prime 2} b)
\nonumber\\
&=
( -2\pi N_\phi (-c X^1+ aX^2) d,\ 2\pi N_\phi (-c X^1+ aX^2) b)
\end{align*}
$U_G(M)$ will change $H'$ to $H$:
\begin{align*}
&\ \ \ \ U_G(M) H' U_G^\dag(M)=H
\nonumber\\
&=
-\sum_k
\frac{1}{2}\sum_{i,j=1,2}
\Big(\frac{\partial}{\partial X^{i}_k} - \imth A_i \Big)
g_{ij}
\Big(\frac{\partial}{\partial X^{j}_k} - \imth A_j \Big)
\end{align*}
with $(A_1,A_2)=(-2\pi N_\phi X^2,0)$.
We find that
\begin{align}
\label{UGuG}
U_G(M) &=\prod_k u_G(M;\v X_k)  ,
\nonumber\\
u_G(M;\v X)&=\e^{\imth  2\pi N_\phi [
bc X^1X^2
-\frac{cd}{2} (X^1)^2
-\frac{ab}{2} (X^2)^2
] }  .
\end{align}

Eqn. \eq{PsiUPsi} can also be rewritten as a transformation
on the wave function $\Phi_{\ga}(\{\v X_i\}|\tau)$:
\begin{align}
\label{UPsi}
U(M)\Phi_{\ga}(\{\v X_i\}|\tau)
&=
U_G(M)\Phi_{\ga}(\{\v X'_i\}|\tau')
\nonumber\\
&=
U_G(M)\Phi_{\ga}(\{M^{-1}\v X_i\}|\tau')
\nonumber\\
&=
\Phi_{\ga'}(\{\v X_i\}|\tau)   U_{\ga'\ga}(M) .
\end{align}
where $\tau'=\frac{M_{12}+M_{22}\tau}{M_{11}+M_{21}\tau}$.
We see that the action of the operator $U(M)$ on a wave function
$\Phi_{\ga}(\{\v X_i\}|\tau) $
is to
replace $\v X_i$ by
$\v X'_i=M^{-1}\v X_i$, replace $\tau$ by $\tau'=\frac{M_{12}+M_{22}\tau}{M_{11}+M_{21}\tau}$,
and then multiply a phase factor $U_G(M)$ given in \eqn{UGuG}.
Thus \eqn{UPsi} defines a way how modular transformations
act on functions.
We find that
\begin{align*}
&\ \ \ \
U(M')\Big( U(M)\Phi_\ga(\{\v X_i\}|\tau) \Big)
\nonumber\\
&= U(M') \Big( U_G(M)\Phi_\ga(\{M^{-1}\v X_i\}|\tau') \Big)
\nonumber\\
&=U_G(M'M) \Phi_\ga(\{M^{-1}M^{\prime -1}\v X_i\}|\tau'')
\nonumber\\
&=U_G(M'M) \Phi_\ga(\{(M'M)^{-1}\v X_i\}|\tau'')
\nonumber\\
&= U(M'M)\Phi_\ga(\{\v X_i\}|\tau)
.
\end{align*}
Here $\tau'=\frac{M_{12}+M_{22}\tau}{M_{11}+M_{21}\tau}$ and
$\tau''=\frac{M_{12}+M_{22}\t\tau}{M_{11}+M_{21}\t\tau}$ with
$\t\tau=\frac{M'_{12}+M'_{22}\tau}{M'_{11}+M'_{21}\tau}$.
Thus
\begin{equation*}
 U(M')U(M)=U(M'M) .
\end{equation*}
So $U(M)$ form a faithful representation of modular transformations $SL(2,Z)$.

To summarize, there are two kinds of deformation loops $\tau(s)$. If
$\tau(0)=\tau(1)$, the deformation loop is contractible [\ie we can
deform the loop to a point, or in other words we can continuously
deform the function $\tau(s)$ to a constant function
$\tau(s)=\tau(0)=\tau(1)$].  For a contractible loop, the associated
non-Abelian Berry's phase is actually a $U(1)$ phase
$U_{\ga\ga'}=\e^{\imth \vphi}\del_{\ga\ga'}$.  where $\vphi$ is path
dependent.  If $\tau(0)$ and $\tau(1)$ are related by a modular
transformation, the deformation loop is non-contractible.  Then the
associated non-Abelian Berry's phase is non-trivial.  If two
non-contractible loops can be deformed into each other continuously,
then the two loops only differ by a contractible loop.  The
associated non-Abelian Berry's phases will only differ by an overall
$U(1)$ phase.  Thus, upto an overall $U(1)$ phase, the non-Abelian
Berry's phases $U_{\ga\ga'}$ of a topologically ordered state are
determined by the modular transformation
$\tau\to\tau'=\frac{a\tau+b}{c\tau+d}$.  We also show that we can
use the parallel transportation to defined a system of basis
$\Phi_\ga(\{\v X_i\}|\tau)$ for all inverse-mass-matrices labeled by
$\tau$. By considering the relation of those basis for two $\tau$'s
related by an modular transformation, we can even obtain a faithful
representation of the modular transformation $SL(2,Z)$.

\subsection{Algebra of modular transformations and translations}

The translation $T_{\v d}$ and modular transformation $U(M)$ all act
within the space of degenerate ground states. There is an algebraic
relation between those operators. From \eq{UPsi}, we see that
\begin{align*}
&\ \ \ \
U(M)T_{\v d} \Phi_\ga(\{\v X_i\}|\tau) = U(M) \Phi_\ga(\{\v X_i+\v d\}|\tau)
\nonumber \\
&=U_G \Phi_\ga(\{M^{-1}\v X_i+\v d\}|\tau')
\nonumber\\
&=U_G \Phi_\ga(\{M^{-1}(\v X_i+M\v d)\}|\tau') .
\end{align*}
Therefore
\begin{equation}
\label{UTTU}
 \e^{\imth \th} U(M)T_{\v d}=T_{M\v d}U(M)  .
\end{equation}

Let us determine the possible phase factor $\e^{\imth \th}$ for some special
cases.
Consider the modular transformation $\tau\to \tau'=\tau+1$ generated by
\begin{equation*}
 M_T=\bpm
1 & 1\\
0 & 1
\epm,\ \ \ \ \
 M_T^{-1}=\bpm
1 & -1\\
0 & 1
\epm .
\end{equation*}
We first calculate
\begin{align*}
&\ \ \ \
U(M_T)T_1 \Phi_\ga(\{\v X_i\}|\tau) = U(M_T) \Phi_\ga(\{\v X_i+\v d_1\}|\tau)
\nonumber \\
&=
\e^{-\imth  \pi N_\phi \sum_i (X_i^2)^2 }
\Phi_\ga(\{M_T^{-1}\v X_i+\v d_1\}|\tau')
.
\end{align*}
where $\v d_1=(\frac{1}{N_\phi},0)$.  We note that
\begin{equation*}
 M_T\v d_1=
\bpm
1 & 1\\
0 & 1
\epm
\bpm
\frac{1}{N_\phi} \\
0
\epm
=\v d_1 .
\end{equation*}
Thus we next calculate
\begin{align*}
&\ \ \ \
T_1 U(M_T) \Phi_\ga(\{\v X_i\}|\tau)
\nonumber\\
&=
T_1\e^{-\imth  \pi N_\phi \sum_i (X_i^2)^2 } \Phi_\ga(\{M_T^{-1}\v X_i\}|\tau')
\nonumber \\
&=
\e^{-\imth  \pi N_\phi \sum_i (X_i^2)^2 }
\Phi_\ga(\{M_T^{-1}(\v X_i+\v d_1)\}|\tau')
\nonumber\\
&=
\e^{-\imth  \pi N_\phi \sum_i (X_i^2)^2 }
\Phi_\ga(\{M_T^{-1}\v X_i+\v d_1\}|\tau')
.
\end{align*}
Therefore
\begin{equation*}
 U(M_T)T_1=T_1U(M_T).
\end{equation*}

To obtain the algebra between $U(M_T)$ and $T_2$, we first calculate
\begin{align*}
&\ \ \ \
U(M_T)T_2 \Phi_\ga(\{\v X_i\}|\tau)
\nonumber\\
&= U(M_T)
\e^{\imth 2\pi \sum_i X^1_i}
\Phi_\ga(\{\v X_i+\v d_2\}|\tau)
\nonumber \\
&=
\e^{-\imth  \pi N_\phi \sum_i (X_i^2)^2 }
\e^{\imth 2\pi \sum_i (X^1_i-X^2_i)}
\Phi_\ga(\{M_T^{-1}\v X_i+\v d_2\}|\tau')
,
\end{align*}
where $\v d_2=(0,\frac{1}{N_\phi})$.
We note that
\begin{equation*}
 M_T\v d_2=
\bpm
1 & 1\\
0 & 1
\epm
\bpm
0\\
\frac{1}{N_\phi} \\
\epm
=\v d_1 +\v d_2.
\end{equation*}
Thus we next calculate
\begin{align*}
&\ \ \ \
T_2T_1U(M_T) \Phi_\ga(\{\v X_i\}|\tau)
\nonumber \\
&=
T_2T_1\e^{-\imth  \pi N_\phi \sum_i (X_i^2)^2 }
\Phi_\ga(\{M_T^{-1}\v X_i\}|\tau')
\nonumber\\
&=
T_2\e^{-\imth  \pi N_\phi \sum_i (X_i^2)^2 }
\Phi_\ga(\{M_T^{-1}(\v X_i+\v d_1)\}|\tau')
\nonumber\\
&=
\e^{\imth 2\pi \sum_i X^1_i}
\e^{-\imth  \pi N_\phi \sum_i (X_i^2+\frac{1}{N_\phi})^2 }  \times
\nonumber\\
&\ \ \ \ \ \ \ \ \ \ \
\Phi_\ga(\{M_T^{-1}(\v X_i+\v d_1+\v d_2)\}|\tau')
\nonumber\\
&=
\e^{-\imth \pi N/N_\phi}
\e^{\imth 2\pi \sum_i (X^1-X^2_i)_i}
\e^{-\imth  \pi N_\phi \sum_i (X_i^2)^2 }  \times
\nonumber\\
&\ \ \ \ \ \ \ \ \ \ \
\Phi_\ga(\{M_T^{-1}\v X_i+\v d_2)\}|\tau')
,
\end{align*}
We see that
\begin{equation*}
 U(M_T)T_2=
\e^{\imth \pi \frac{n}{m}}
T_2T_1U(M_T).
\end{equation*}

Next we consider the modular transformation $\tau\to \tau'=-1/\tau$
generated by
\begin{equation*}
 M_S=\bpm
0 & -1\\
1 & 0
\epm,\ \ \ \ \
 M_S^{-1}=\bpm
0 & 1\\
-1 & 0
\epm .
\end{equation*}
From
\begin{align*}
U(M_S) \Phi_\ga(\{\v X_i\}|\tau)
&=
\e^{-\imth  2 \pi N_\phi \sum_i X_i^1 X_i^2 }
\Phi_\ga(\{M_S^{-1}\v X_i\}|\tau')
,
\end{align*}
we find
\begin{align*}
&\ \ \ \
U(M_S) U(M_S) \Phi_\ga(\{\v X_i\}|\tau)
\nonumber\\
&= U(M_S)
\e^{-\imth  2 \pi N_\phi \sum_i X_i^1 X_i^2 }
\Phi_\ga(\{M_S^{-1}\v X_i\}|\tau')
\nonumber\\
&=
\e^{-\imth  2 \pi N_\phi \sum_i X_i^1 X_i^2 }
\e^{-\imth  2 \pi N_\phi \sum_i X_i^2 (-X_i^1) }
\Phi_\ga(\{-\v X_i\}|\tau)
\nonumber\\
&=
\Phi_\ga(\{-\v X_i\}|\tau)
.
\end{align*}
We see that $U(M_S) U(M_S)=U(-1)$ generates to transformation $\v X\to -\v X$.
We can show that
\begin{equation}
\label{Um1T12}
U(-1) T_1 U^{-1}(-1)=T^{-1}_1,\ \ \ \ \ \
U(-1) T_2 U^{-1}(-1)=T^{-1}_2 .
\end{equation}

Since the wave function of an orbital satisfies
\begin{equation*}
 \phi^{(l)}(-X^1,-X^2)=
 \phi^{(N_\phi-l)}(X^1,X^2)=
 \phi^{(-l)}(X^1,X^2),
\end{equation*}
we find
\begin{align}
\label{CPhi}
U(M_S) U(M_S) \Phi_\ga(\{\v X_i\}|\tau)
&=
U(-1) \Phi_\ga(\{\v X_i\}|\tau)
\nonumber\\
&=
\Phi_{\ga^*}(\{\v X_i\}|\tau),
\end{align}
where $\ga^*$ corresponds to
the occupation distribution
\begin{equation*}
 n_{\ga^*;l+l_s}=n_{\ga;-l+l_s} .
\end{equation*}
Since $n_{\ga,l}$ and $n_{\ga^*,l}$ are periodic with a period of
$m$, the above can be rewritten as
\begin{equation}
\label{nbarganga}
 n_{\ga^*;l}=n_{\ga;(l_{max}-l)\% m}
.
\end{equation}
$\ga^*$ also corresponds to the anti-quasiparticle of $\ga$.
Thus we find that
\begin{equation*}
 [U(M_S)]^2=C,\ \ \ \ \ \
C\Phi_\ga=\Phi_{\ga^*}
  ,
\end{equation*}
where $C$ is the quasiparticle conjugation operator.
Clearly $C^2=1$.

Let us first calculate
\begin{align*}
&\ \ \ \
U(M_S) T_1 \Phi_\ga(\{\v X_i\}|\tau)
\nonumber\\
&= U(M_S)
\Phi_\ga(\{M_S^{-1}\v X_i+\v d_1\}|\tau)
\nonumber\\
&=
\e^{-\imth  2 \pi N_\phi \sum_i X_i^1 X_i^2 }
\Phi_\ga(\{M_S^{-1}\v X_i+\v d_1\}|\tau')
.
\end{align*}
Since $M_S\v d_1=\v d_2$, we next consider
\begin{align*}
&\ \ \ \
T_2 U(M_S)  \Phi_\ga(\{\v X_i\}|\tau)
\nonumber\\
&=
T_2 \e^{-\imth  2 \pi N_\phi \sum_i X_i^1 X_i^2 }
\Phi_\ga(\{M_S^{-1}\v X_i\}|\tau')
\nonumber\\
&=
\e^{\imth 2\pi \sum_i X^1_i}
\e^{-\imth  2 \pi N_\phi \sum_i X_i^1 (X_i^2+\frac{1}{N_\phi}) } \times
\nonumber\\
&\ \ \ \ \ \ \ \ \ \ \ \
\Phi_\ga(\{M_S^{-1}(\v X_i+\v d_2)\}|\tau')
\nonumber\\
&=
\e^{-\imth  2 \pi N_\phi \sum_i X_i^1 X_i^2 }
\Phi_\ga(\{M_S^{-1}(\v X_i+\v d_2)\}|\tau')
.
\end{align*}
We see that
\begin{align*}
 U(M_S)T_1&=T_2 U(M_S) ,
\nonumber\\
 U(M_S)T_2&=T^{-1}_1 U(M_S) ,
\end{align*}
where we have used \eq{Um1T12}.
Let us introduce
\begin{equation*}
 T=\e^{\imth \th} U(M_T),\ \ \ \ \ \
 S= U(M_S),
\end{equation*}
where the value of $\th$ will be chosen to make $T_{00}=1$.
We find that
\begin{align}
\label{TST1T2a}
 TT_1&=T_1T, &  TT_2&=\e^{\imth \pi\frac{n}{m}} T_2T_1 T,
\nonumber\\
 ST_1&=T_2S, &  ST_2&= CT_1 C^{-1} S.
\end{align}

\subsection{Relation to conformal field theory}
\label{cftsec}

The symmetric polynomial $\Phi$ can be written as a correlation function of
vertex operators $V_e(z)$ in a conformal field theory
(CFT):\cite{MR9162,WWopa,WWHopa}
\begin{equation}
\label{PhiVe}
 \Phi(\{z_i\})=\lim_{z_\infty\to \infty} z_\infty^{2h_N}
\<V(z_\infty)\prod_i V_e(z_i) \>
\end{equation}
$V_e$ (which will be called an electron operator) has a form
\begin{equation*}
 V_e(z)=\psi(z)\e^{\imth \phi(z)/\sqrt{\nu}}
\end{equation*}
where $\psi$ is a simple current operator and $\e^{\imth \phi(z)/\sqrt{\nu}}$
is the vertex operator on a Gaussian model with a scaling dimension
$h=\frac{1}{2\nu}$.  The scaling dimension $h^\text{sc}_a$ of $\psi_a(z)\equiv
[\psi(z)]^a$ has being calculated from the pattern of zeros $\{S_a\}$ in
\Ref{WWsymm}:
\begin{align}
\label{hscaSa}
 h^\text{sc}_a = S_a- \frac{aS_n}{n}+\frac{am}{2}-\frac{a^2m}{2n}.
\end{align}

The quasiparticle state $\Phi_\ga$ can also be expressed as
 a correlation function in a CFT:
\begin{equation}
\label{PhiVeq}
 \Phi_\ga(\{z_i\})=\lim_{z_\infty\to \infty} z_\infty^{2h^q_N}
\<V_q(z_\infty)V_\ga(0)\prod_i V_e(z_i) \>
\end{equation}
(Note that $\Phi_\ga$ has a quasiparticle at $z=0$.)
Here $V_\ga$ is a quasiparticle operator in CFT which has a form
\begin{equation}
\label{Vga}
 V_\ga(z)= \si_\ga(z) \e^{\imth \phi(z) Q_\ga/\sqrt{\nu}}
\end{equation}
where $\si_\ga(z)$ is a ``disorder'' operator in the CFT generated
by the simple current operator $\psi$.  Different quasiparticles
labeled by different $\ga$ will correspond to different ``disorder''
operators.

Let us introduce a quantitative way to characterize the quasiparticle operator.
We first fuse the quasiparticle operator with $a$ electron operators:
\begin{align}
\label{Vgaa}
 V_{\ga;a}(z) &=V_\ga V_a = \si_{\ga;a}(z) \e^{\imth \phi(z) Q_{\ga;a}/\sqrt{\nu}}
\nonumber\\
 \si_{\ga;a}&= \si_{\ga}\psi_a,\ \ \ \ Q_{\ga;a}=Q_{\ga}+a ,
\end{align}
where
$V_a \equiv (V_e)^a=\psi_a \e^{\imth a \phi(z)a/\sqrt{\nu}} $.
Then, we consider the operator product expansion (OPE) of $V_{\ga;a}$ with
$V_e$
\begin{equation}
\label{VeVga}
 V_e(z) V_{\ga;a}(w)=(z-w)^{l^\text{CFT}_{\ga;a+1}} V_{\ga;a+1}(w) .
\end{equation}
Let $h_a$, $h_\ga$, and $h_{\ga;a}$ be the scaling dimensions
of $V_a$, $V_\ga$, and $V_{\ga;a}$ respectively. We have
\begin{equation}
\label{lhhh}
 l^\text{CFT}_{\ga;a+1}=h_{\ga;a+1}-h_1-h_{\ga;a} .
\end{equation}
Since the quasiparticle wave function $\Phi_\ga(\{z_i\})$ must be a single
valued function of $z_i$'s, this requires that
$l^\text{CFT}_{\ga;a}$ must be integers.
The sequence of integers $\{l^\text{CFT}_{\ga;a}\}$ gives us a quantitative
way to characterize quasiparticle operators $V_\ga$
in CFT.

From the occupation distribution description of the quasiparticle $\ga$
introduced in section \ref{qppoz}, we see that a quasiparticle can also be
characterized by another sequence of integers $\{l_{\ga;a}\}$.  What is the
relation between the two sequences of integers, $\{l^\text{CFT}_{\ga;a}\}$ and
$\{l_{\ga;a}\}$, that characterize the same set of quasiparticles.  From
\eqn{VeVga}, we see that $l^\text{CFT}_{\ga;a}$ is the order of zeros as we
move an electron $z_i$ towards a quasiparticle $\ga$ fused with $a$-electrons.
Thus $l^\text{CFT}_{\ga;a}$ is the order of zero $D_{\ga;a,1}$ introduced in
section \ref{qPOZ}. From \eq{DgaSS} and $S_1=0$, we find that
$l^\text{CFT}_{\ga;a}$ in the above OPE is given by
$l^\text{CFT}_{\ga;a}=S_{\ga,a}-S_{\ga,a-1}$.  Thus the two sequences,
$\{l^\text{CFT}_{\ga;a}\}$ and $\{l_{\ga;a}\}$, are identical
$\{l^\text{CFT}_{\ga;a}\} = \{l_{\ga;a}\}$.  In the rest of the paper, we will
drop the superscript CFT in $l^\text{CFT}_{\ga;a}$.

Now let us calculate the quasiparticle charge $Q_\ga$ (see \eqn{Vga}) from the
sequence $\{l_{\ga;a}\}$ within the CFT.  Using
$l_{\ga;a}=S_{\ga,a}-S_{\ga,a-1}$, we can rewrite \eq{lhhh} as
$ h_{\ga;a+1}-h_{\ga;a} =S_{\ga;a+1}-S_{\ga;a} +h_1 $.
Thus
$ h_{\ga;a} - h_\ga = S_{\ga;a} +ah_1 $,
where we have used $S_{\ga;0}=S_\ga=0$.
Using $h_1=h^\text{sc}_1+\frac{1}{2\nu}=\frac{m}{2}-\frac{S_n}{n}$,
we find
\begin{equation*}
 h_{\ga;a} - h_\ga = S_{\ga;a} +a\Big( \frac{m}{2}-\frac{S_n}{n}  \Big).
\end{equation*}
Since $\si_{\ga,n}=\si_\ga$, we have
\begin{equation*}
 h_{\ga;n} - h_\ga = \frac{(Q_\ga+n)^2-Q_\ga^2}{2\nu}
=S_{\ga;n} +n\Big( \frac{m}{2}-\frac{S_n}{n}  \Big)
.
\end{equation*}
Thus
\begin{align*}
 Q_\ga=\frac{S_{\ga;n}-S_n}{m}=\frac{1}{m}\sum_{a=1}^n (l_{\ga;a}-l_a)  ,
\end{align*}
which agrees with \eqn{Qgala}.

Let $h^\text{sc}_{\ga;a}$ be the scaling dimension of $\si_\ga\psi_a$.  We see
that $h^\text{sc}_{\ga;a} = h_{\ga;a}-\frac{(Q_\ga+a)^2}{2\nu}$ and
\begin{equation*}
 h^\text{sc}_{\ga;a} -h^\text{sc}_\ga
= S_{\ga;a} -\frac{m }{2n}a^2+\Big(\frac{m}{2}-\frac{S_{\ga;n}}{n}\Big)a
\end{equation*}
Let
\begin{equation*}
 s_{\ga;a}=S_{\ga;a} -\frac{m}{2n}a^2
+\Big(\frac{m}{2}-\frac{S_{\ga;n}}{n}\Big)a
=s_{\ga;a+n}
\end{equation*}
we can rewrite the above as
\begin{equation}
\label{hgasga}
  h^\text{sc}_{\ga;a} = h^\text{sc}_\ga +s_{\ga;a} .
\end{equation}
We see that the simple-current part of CFT is determined by $s_{\ga;a}$,
$a=1,\cdots,n-1$, only.
In particular, if
$\si_{\ga'}$
and $\si_{\ga}$ are related by a simple current operator,
$\si_{\ga'}=\si_{\ga} \psi_a$, then the scaling
dimension of $\si_{\ga'}$ can be calculated from that
of $\si_{\ga}$:
$ h^\text{sc}_{\ga'} = h^\text{sc}_{\ga} + s_{\ga;a} $.


\end{document}